\documentclass[twocolumn,trackchanges]{aastex62}

\usepackage{amsmath}
\usepackage{multirow}
\usepackage{color}
\usepackage{graphicx}

\newcommand{\hii}{\ion{H}{2}}

\newcommand{\gof}{$G_{\rm 0}$}

\newcommand{\cii}{\mbox{[\ion{C}{2}]}}

\newcommand{\ois}{\mbox{[\ion{O}{1}]}}
\newcommand{\oi}{\mbox{[\ion{O}{1}]\,63\,$\mu$m}}
\newcommand{\oitab}{\mbox{[\ion{O}{1}]}}
\newcommand{\oii}{\mbox{[\ion{O}{1}]\,145\,$\mu$m}}
\newcommand{\oiitab}{\mbox{[\ion{O}{1}]}}
\newcommand{\oiii}{\mbox{[\ion{O}{3}]\,88\,$\mu$m}}
\newcommand{\oiiitab}{\mbox{[\ion{O}{3}]}}

\newcommand{\nit}{[\ion{N}{2}]}
\newcommand{\nii}{[\ion{N}{2}]\,122\,$\mu$m}
\newcommand{\niitab}{[\ion{N}{2}]}

\newcommand{\niii}{[\ion{N}{3}]\,57\,$\mu$m}
\newcommand{\niiitab}{[\ion{N}{3}]}

\newcommand{\lfir}{$L_{\rm FIR}$}
\newcommand{\sfir}{$\Sigma_{\rm FIR}$}
\newcommand{\firmol}{$L_{\rm FIR}/M_{\rm mol}$}
\newcommand{\sblue}{$S_{63\,\mu\mathrm{m}}$}
\newcommand{\sgreen}{$S_{100\,\mu\mathrm{m}}$}
\newcommand{\sred}{$S_{122\,\mu\mathrm{m}}$}
\newcommand{\sratio}{$S_{63\,\mu\mathrm{m}}/S_{122\,\mu\mathrm{m}}$}
\newcommand{\jone}{\mbox{J=1--0}}

\shorttitle{SHINING, A Survey of Far Infrared Lines in Nearby Galaxies}
\shortauthors{Herrera-Camus et al.}

\begin{document}

\title{SHINING, A Survey of Far Infrared Lines in Nearby Galaxies. I: Survey Description, Observational Trends, and Line Diagnostics}

\correspondingauthor{R. Herrera-Camus}
\email{rhc@mpe.mpg.de}

\author{R. Herrera-Camus}
\affil{Max-Planck-Institut f\"{u}r Extraterrestrische Physik (MPE), Giessenbachstr., D-85748 Garching, Germany}

\author{E. Sturm}
\affiliation{Max-Planck-Institut f\"{u}r Extraterrestrische Physik (MPE), Giessenbachstr., D-85748 Garching, Germany}

\author{J. Graci\'a-Carpio}
\affiliation{Max-Planck-Institut f\"{u}r Extraterrestrische Physik (MPE), Giessenbachstr., D-85748 Garching, Germany}

\author{D. Lutz}
\affiliation{Max-Planck-Institut f\"{u}r Extraterrestrische Physik (MPE), Giessenbachstr., D-85748 Garching, Germany}

\author{A. Contursi}
\affiliation{Max-Planck-Institut f\"{u}r Extraterrestrische Physik (MPE), Giessenbachstr., D-85748 Garching, Germany}

\author{S. Veilleux}
\affiliation{Department of Astronomy and Joint Space-Science Institute, University of Maryland, College Park, MD 20742, USA}

\author{J. Fischer}
\affiliation{Naval Research Laboratory, Remote Sensing Division, 4555 Overlook Avenue SW, Washington DC 20375, USA}

\author{E. Gonz\'alez-Alfonso}
\affiliation{Departamento de F\'isica y Matem\'aticas, Universidad de Alcal\'a, Campus Universitario, E-28871 Alcal\'a de Henares, Madrid, Spain}

\author{A. Poglitsch}
\affiliation{Max-Planck-Institut f\"{u}r Extraterrestrische Physik (MPE), Giessenbachstr., D-85748 Garching, Germany}

\author{L. Tacconi}
\affiliation{Max-Planck-Institut f\"{u}r Extraterrestrische Physik (MPE), Giessenbachstr., D-85748 Garching, Germany}

\author{R. Genzel}
\affiliation{Max-Planck-Institut f\"{u}r Extraterrestrische Physik (MPE), Giessenbachstr., D-85748 Garching, Germany}

\author{R. Maiolino}
\affiliation{Kavli Institute for Cosmology, University of Cambridge, Madingley Road, Cambridge CB3 0HA, UK}

\author{A. Sternberg}
\affiliation{Raymond and Beverly Sackler School of Physics \& Astronomy, Tel Aviv University, Ramat Aviv 69978, Israel}

\author{R. Davies}
\affiliation{Max-Planck-Institut f\"{u}r Extraterrestrische Physik (MPE), Giessenbachstr., D-85748 Garching, Germany}

\author{A. Verma}
\affiliation{Oxford University, Dept. of Astrophysics, Oxford OX1 3RH, UK}



\begin{abstract}
We use the Herschel/PACS spectrometer to study the global and spatially resolved far-infrared (FIR) fine-structure line emission in a sample of 52 galaxies that constitute the SHINING survey. These galaxies include star-forming, active-galactic nuclei (AGN), and luminous infrared galaxies (LIRGs). We find an increasing number of galaxies (and kiloparsec size regions within galaxies) with low line-to-FIR continuum ratios as a function of increasing FIR luminosity (\lfir), dust infrared color, \lfir\ to molecular gas mass ratio (\firmol), and FIR surface brightness (\sfir). The correlations between the \cii/FIR or \ois/FIR ratios with \sfir\ are remarkably tight ($\sim0.3$~dex scatter over almost four orders of magnitude in \sfir). We observe that galaxies with $L_{\mathrm{FIR}}/M_{\mathrm{mol}} \gtrsim 80\,L_{\odot}\,M_{\odot}^{-1}$ and $\Sigma_{\rm FIR}\gtrsim10^{11}$~$L_{\odot}$~kpc$^{-2}$ tend to have weak fine-structure line-to-FIR continuum ratios, and that LIRGs with infrared sizes $\gtrsim1$~kpc have line-to-FIR ratios comparable to those observed in typical star-forming galaxies. We analyze the physical mechanisms driving these trends in Paper II \citep{rhc_rhc18b}. The combined analysis of the \cii, \nii, and \oiii\ lines reveals that the fraction of the \cii\ line emission that arises from neutral gas increases from 60\% to 90\% in the most active star-forming regions and that the emission originating in the ionized gas is associated with low-ionization, diffuse gas rather than with dense gas in \hii\ regions. Finally, we report the global and spatially resolved line fluxes of the SHINING galaxies to enable the comparison and planning of future local and high-$z$ studies.
\end{abstract}




\section{Introduction} \label{sec:intro}

The {\it Herschel Space Observatory} \citep{rhc_pilbratt10} has transformed our view of the far-infrared (FIR) universe. In addition to significantly higher sensitivity and spectral resolution, one of the major steps forward achieved by {\it Herschel} was to enable observations of the dust continuum and the fine-structure transitions in the $\sim55-500~\mu$m range with spatial resolution at least 4 times higher than previous infrared space missions. For nearby galaxies, {\it Herschel} made possible for the first time to resolve and study a large number of diverse galactic environments. In this context, the {\it Herschel} guaranteed time program ``Survey with Herschel of the Interstellar medium in Nearby INfrared Galaxies'' (SHINING; PI Sturm) was planned to obtain a comprehensive view of the physical processes at work in the interstellar medium of local galaxies, ranging from objects with moderately enhanced star formation and nuclear activity to the most dense, energetic, and obscured environments in ultra-luminous infrared galaxies.

Each of the SHINING galaxies was observed in the six brightest photodissociation region (PDR) and \hii\ region lines in the {\it Herschel}/PACS \citep{rhc_poglitisch10} spectral range; namely the \cii~158~$\mu$m, \oii, and \oi\ PDR lines, and the \nii, \oiii, and \niii\ \hii\ lines (see Table~\ref{table:lines} for details). Among these, the \cii~158~$\mu$m is commonly the brightest and the \cii\ line flux in typical, star-forming galaxies is $\sim4000$ brighter than the CO~\jone\ line \citep{rhc_wolfire89,rhc_hughes16b}. With an ionization potential of only 11.2~eV, \cii\ emission can arise from both ionized and neutral gas phases. For the latter, \cii\ is the main coolant \citep{rhc_wolfire03}, and for a system in thermal equilibrium, the \cii\ emission is a measure of the heating rate. This heating is, in normal galaxies, dominated by FUV radiation processed by dust through the photoelectric heating effect, and is thus a consequence of star formation activity. Hence, a connection between the \cii\ emission and the star formation rate is expected \citep[e.g.,][]{rhc_stacey91,rhc_boselli02,rhc_delooze14,rhc_rhc15,rhc_nordon16}. 

Together with \cii, the other main cooling channel for the neutral, atomic gas is the \oi\ line \citep{rhc_wolfire03}. In fact, due to its higher excitation temperature and critical density the \oi\ line can become the major coolant in warm, dense gas environments such as those found in starburst regions or the central parts of galaxies \citep[e.g., ][]{rhc_brauher08,rhc_croxall12}. The other neutral oxygen fine-structure transition is \oii, which is typically a factor $\sim10$ weaker than the \oi\ line \citep{rhc_brauher08,rhc_vasta10}, but is less likely to suffer absorption.

Finally, the three ionic FIR fine-structure lines included in our survey are \niii, \oiii, and \nii. The first two are closely associated with dense \hii\ regions as ionization energies of 35.1~eV and 47.5~eV are required to create O$^{++}$ and N$^{++}$ ions, respectively. The \nii\ line, on the other hand, can also arise from lower excitation gas with typical electron densities in the Milky~Way and star-forming galaxies of about $n_{\rm e}\approx30$~cm$^{-3}$ \citep{rhc_goldsmith15,rhc_rhc16}. 

The fine-structure FIR lines included in our study span a wide range of ionization potentials ($\chi\sim11-35$~eV), critical densities ($n_{\rm crit}\sim40-10^6$~cm$^{-3}$), and excitation temperatures ($T_{\rm ex}\sim90-300$~K). Together they are useful probes of the physical conditions of the interstellar medium (ISM), including radiation fields, gas densities, temperatures, and abundances \citep[e.g.,][]{rhc_rubin85,rhc_tielens85,rhc_sternberg89,rhc_sternberg95,rhc_kaufman99,rhc_kaufman06,rhc_fischer14}. For example, the relative intensity of the cooling lines of the neutral gas (mainly \cii\ and \ois) compared to the FIR continuum yields a measure of the gas heating efficiency. Over the last two decades a number of studies have shown that the (\cii+\ois)/FIR ratio decreases as a function of increasing FIR luminosity (\lfir), FIR color (e.g., 60~$\mu$m/100~$\mu$m), and FIR surface brightness (\sfir) \citep[e.g.,][]{rhc_malhotra97,rhc_malhotra01,rhc_brauher08,rhc_gracia-carpio11,rhc_diaz-santos13,rhc_diaz-santos17,rhc_croxall13,rhc_ibar15,rhc_lutz16,rhc_smith17}. The proposed explanations for the decrease in the (\cii+\ois)/FIR continuum ratios are multiple, including a reduction of the photoelectric yield due to the charging and/or destruction of dust grains \citep[e.g.,][]{rhc_malhotra97,rhc_croxall12}, absorption of UV photons by dust in \hii\ regions \citep[e.g.,][]{rhc_abel09,rhc_gracia-carpio11}, line self-absorption \citep[e.g.,][]{rhc_gonzalez-alfonso14}, softer UV radiation coming from older stellar populations \citep[e.g.,][]{rhc_malhotra01,rhc_smith17}, the impact on the ionization state of the gas by AGNs \citep[e.g.,][]{rhc_langer15}, among others. In this paper we focus on presenting the trends of the line-to-FIR continuum ratios as a function of different galaxy properties, leaving the detailed analysis of the physical mechanisms responsible for the line deficits for Paper II \citep{rhc_rhc18b}. In the particular case of the \cii\ line, in Paper II we find that the decrease in the \cii/FIR ratio as function of FIR surface brightness can be explained as a combination of a reduction in the dust photoelectric heating efficiency, an increase in the ionization parameter, and the inability of the \cii\ line to track the increase in the FUV radiation field as galaxies become more compact and luminous.

There are many other examples of line diagnostics in the far-infrared that have been used to characterize the conditions in the ISM. For example, the \nit/\cii\ line ratio serves as a tracer of the fraction of the \cii\ line emission that arises from ionized gas. On nearby galaxies and LIRGs, this fraction have been found to decrease from $\sim30\%$ to $\sim5\%$ in the most active, star-forming environments \citep[e.g.,][]{rhc_croxall17,rhc_diaz-santos17}. Other example is the \oiii/\nii\ line ratio, which provides a sensitive probe of the UV field hardness and has the advantage to be practically insensitive to the gas density \citep[e.g.,][]{rhc_rubin85,rhc_ferkinhoff11}. Finally, in this paper we discuss a new diagnostic diagram based on the \cii/\oi\ (sensitive to the gas density and radiation field strength) and \niii/\nii\ (sensitive to the hardness of the UV radiation field) line ratios that can help to discriminate between star-formation and AGN activity on $\sim$kiloparsec scales.

The aim of this paper is to introduce the SHINING sample and present the main results of our analysis of fine-structure atomic and ionic transitions. Other works based on the SHINING dataset include studies of galaxy outflows \citep{rhc_sturm11,rhc_contursi13,rhc_gonzalez-alfonso14,rhc_gonzalez-alfonso17,rhc_janssen16}, ISM diagnostics \citep{rhc_fischer10,rhc_gracia-carpio11,rhc_gonzalez-alfonso12,rhc_gonzalez-alfonso15,rhc_contursi17}, and molecular gas excitation \citep{rhc_hailey-dunsheath12,rhc_gonzalez-alfonso13,rhc_mashian15}. The structure of the paper is the following. In Section~\ref{sample} we introduce the SHINING sample of galaxies. In Section~\ref{observations} we describe the {\it Herschel} observations. In Section~\ref{data:reduction} we discuss how the data were reduced and fluxes were measured. In Section~\ref{sec:results} we discuss how the relative intensity of the infrared fine-structure lines (line-to-continuum and line-to-line ratios) on kiloparsec and global scales varies as a function of AGN activity, FIR surface brightness, and star formation efficiency. We present our summary and conclusions in Section~\ref{conclusions}.

\floattable
\begin{deluxetable}{lcccccc}
\tabletypesize{\footnotesize} 
\tablecaption{Fine-structure Lines in the SHINING Survey \label{table:lines}}
\tablehead{
\colhead{Line} & \colhead{Configuration} & \colhead{$\lambda$} & \colhead{$E_{\rm ion}$} & \colhead{$T_{\rm ex}$} & \colhead{$n_{\rm crit,H}~(T=100~K)$} & \colhead{$n_{\rm crit,e}~(T=10^4~K)$} \\ 
\colhead{} & \colhead{} & \colhead{$\mu$m} & \colhead{eV} & \colhead{K} & \colhead{cm$^{-3}$} & \colhead{cm$^{-3}$}}
\startdata
\vspace{-0.1cm}
\cii\ & $^{2}{\rm P}_{3/2}-^{2}{\rm P}_{1/2}$ & 157.74  &  11.26 &   91  & $2.0\times10^3$ & 47 \\ \vspace{-0.1cm}
\oiitab\ & $^{3}{\rm P}_{1}-^{3}{\rm P}_{0}$ & 145.53 & $-$ &   326  &  $2.3\times10^4$ & $-$ \\
\niitab\ & $^{3}{\rm P}_{2}-^{3}{\rm P}_{1}$ & 121.90  &  14.53 &   188  & $-$ & 375 \\ 
\oiiitab\ & $^{3}{\rm P}_{1}-^{3}{\rm P}_{0}$ & 88.36 &  35.12 &  163  &  $-$ &  510 \\
\oitab\ & $^{3}{\rm P}_{2}-^{3}{\rm P}_{1}$ & 63.19 &  $-$ &   227  &  $2.5\times10^5$ & $-$ \\
\niiitab\ & $^{3}{\rm P}_{3/2}-^{3}{\rm P}_{1/2}$ & 57.34 &  29.60 &  251  & $-$ & $3.0\times10^3$ \\
\enddata
\tablecomments{The information in Table~\ref{table:lines} are based on \cite{rhc_draine_book} and the Leiden Atomic and Molecular Database \citep{rhc_schoeier05}.}
\end{deluxetable}

\section{Galaxy sample}\label{sample}

The SHINING sample consists of 52 galaxies in the local Universe ($z < 0.2$) observed with the PACS spectrometer \citep{rhc_poglitisch10} on board {\it Herschel} \citep{rhc_pilbratt10} in multiple atomic and molecular transitions. SHINING was one of the {\it Herschel} Guaranteed Time Observations (GTO) programs, and also included observations of 50 low-metallicity galaxies that constitute the ``Dwarf Galaxy Survey'' \citep{rhc_madden13} which we do not include in our analysis.

The main properties of the SHINING galaxy sample are listed in Table~\ref{table:SHINING sample}. Due to the limited amount of time available for the SHINING GTO observations and the pioneering character of the study, there was no specific criteria involved in the selection of the starburst and Seyfert galaxies rather than choosing well known or archetypical systems (i.e., IR bright, template systems with a wealth of ancillary data available). The \hii/Starburst galaxies are all part of the IRAS Revised Bright Galaxy Sample \citep[RGBS, ][]{rhc_sanders03} and have been observed with {\it ISO} and {\it Spitzer}. The {\it Seyfert~1 AGN} are mostly taken from the CfA Catalogue \citep{rhc_huchra92}, with 3 additional bright objects from \cite{rhc_nandra97}.  We also added the {\it Circinus} galaxy, which is not part of either of the above galaxy samples, but well known and extensively studied at many wavelengths. The {\it Seyfert 2 AGN} are taken from the \cite{rhc_bassani99} sample. The local (U)LIRG ($z<0.1$) sample consists of all ULIRGs in the RBGS (IRAS~F09111--1007 was however not observed due to time limitations). For all galaxy types, additional spectral classifications are based mostly on the Catalogue of Quasars and Active Nuclei \citep[][]{rhc_cetty06} and the catalogue of optical spectroscopy of (U)LIRGs \citep{rhc_veilleux95,rhc_veilleux99}. In order to simplify the comparison between starburst and AGN, we treat composite AGN/starburst systems such as NGC~1365 \citep[e.g.,][]{rhc_turner93,rhc_kristen97} and NGC~4945 \citep{rhc_schurch02} as AGN, and Seyfert types 1.2, 1.5, 1.8, and 1.9 as Seyfert~1 objects.

Table~\ref{table:SHINING sample} also includes two galaxies that were not initially part of SHINING, but have been added to the analysis in this paper because they share similar properties and observations with the original sample: NGC~3079, a Seyfert~2 galaxy observed under the director's discretionary time proposal DDT\_esturm\_4, and IRAS~15206+3342, a ULIRG observed during the {\it Herschel} performance verification (PV) phase. Table~\ref{table:SHINING sample} also includes the FIR surface brightness of our galaxies as reported by \cite{rhc_lutz16}. Briefly, these surface brightnesses are based on the FIR luminosity and a FIR galaxy size derived from a 2-dimensional Gaussian fit to the 70~$\mu$m image of the galaxy, with PSF width subtracted in quadrature.

We include in our analysis molecular gas masses ($M_{\rm mol}$). Following \cite{rhc_gracia-carpio11}, these were measured converting previously published CO observations \citep{rhc_aalto95,rhc_albrecht04,rhc_albrecht07,rhc_andreani95,rhc_baan08,rhc_braun11,rhc_chung09,rhc_combes07,rhc_combes11,rhc_elfhag96,rhc_evans05,rhc_evans09,rhc_gao01,rhc_gao04,rhc_garcia-burillo12,rhc_heckman89,rhc_maiolino97,rhc_mirabel90,rhc_sanders91,rhc_solomon97,rhc_strong04,rhc_tacconi10,rhc_vollmer08,rhc_xia12,rhc_yao03,rhc_young95,rhc_young11,rhc_demello02} using an $\alpha_{\rm CO}$ conversion factor that varies with IRAS $S_{60\mu {\rm m}}/S_{100\mu {\rm m}}$ color in order to reproduce ULIRG ($\approx0.8~M_{\odot}~(K~km~s^{-1}~pc^{2})^{-1}$) and Milky-Way ($\approx4~M_{\odot}~(K~km~s^{-1}~pc^{2})^{-1}$) adequate values \citep{rhc_bolatto13}.

\section{Observations}\label{observations}

PACS provided the first opportunity to obtain high spatial resolution spectral images of galaxies in the far-infrared range between 55~$\mu$m and 210~$\mu$m \citep{rhc_poglitisch10}. The PACS integral field spectrometer consists of 5$\times$5 spatial pixels covering a field of view (FOV) of 47\arcsec$\times$47\arcsec. Each spatial pixel is connected to two photoconductor arrays of 16 spectral pixels, that scan in time the desired (red and blue) wavelength ranges with the help of a grating. The instrumental velocity resolution depends on the observed wavelength range and the selected grating order, but is typically in the range $100-330$~km~s$^{-1}$ when used in 1st or 2nd order, enough to study line velocity profiles from external galaxies.

Most of the SHINING observations were done using the PACS line and range spectroscopy astronomical observing templates (AOTs) in combination with the PACS pointed observing mode. Background subtraction was achieved with the chopping/nodding technique. The chopper throw was varied from 6\arcmin\ for the closest galaxies, to 1\arcmin\ for the most compact systems. In the pointed observing mode, the PACS spectrometer camera observed always at the same chopper ON position, giving by the end of the observation an individual line and continuum measurement in each of the 25 spatial pixels. This was the recommended mode to use for the detection of weak lines in galaxies that are smaller than the maximum chopper throw (6\arcmin). The data were obtained in short time integrations ($\sim$1\,hr) spread over a period of almost two years, from November 2009 to the end of September 2011. 

\startlongtable
\begin{deluxetable*}{lccccc}
\tabletypesize{\footnotesize} 
\tablecaption{SHINING Galaxy Sample \label{table:SHINING sample}}
\tablehead{
\colhead{Source} & \colhead{Optical} & \colhead{Redshift\tablenotemark{a}} & \colhead{$D_{\rm L}$\tablenotemark{a}} & \colhead{log$_{10}$(\sfir)} & \colhead{log$_{10}(L_{\rm{FIR}})$}\\ 
\colhead{} & \colhead{class.} & \colhead{} & \colhead{Mpc} & \colhead{$L_{\odot}$~kpc$^{-2}$} & \colhead{$L_{\odot}$}}
\startdata  
\vspace{0.1cm}
{\it Starburst/Star-forming galaxies}   &    &   &   &  &  \\ \hline
NGC 253                          & HII    &  0.00081   &     3.42  &  12.15 & 10.35 \\ \vspace{-0.1cm}
M 82                             	& HII    &  0.00067 &     3.83  &  \ldots & 10.67 \\ \vspace{-0.1cm}
M 83                             	& HII    &  0.00171 &     7.24  &  \ldots & 10.13 \\ \vspace{-0.1cm}
NGC 1808                         & HII    &  0.00331 &    14.02  &  10.63 & 10.51 \\ \vspace{-0.1cm}
NGC 7552                         & HII    &  0.00536 &    22.74  &  10.90 & 10.79 \\ \vspace{-0.1cm}
Antennae                         & HII    &  0.00568 &    24.10  &  \ldots & 10.74 \\ \vspace{-0.1cm}
NGC 3256                         & HII    &  0.00935 &    39.80  &  10.79 & 11.62 \\ \vspace{-0.1cm}  
IRAS 23128-5919                 & HII &  0.04460 &   195.05  &  10.22 & 11.92 \\ \hline \vspace{0.1cm}
{\it Seyfert galaxies}   &    &   &   &  &  \\ \hline
Circinus                         	& Sy2    &  0.00145 &     6.13  &  11.49 &  9.92 \\ \vspace{-0.1cm}
Cen A                            & Sy2    &  0.00182 &     7.70  & 9.19  &  9.91 \\ \vspace{-0.1cm}
NGC 4945\tablenotemark{b}   & Sy2    &  0.00187 &     7.91  &  11.85 & 10.28 \\ \vspace{-0.1cm}
NGC 4051                         & Sy1    &  0.00233 &    9.86  &  10.93 &  9.84 \\ \vspace{-0.1cm}
NGC 1386                         & Sy2    &  0.00289 &    12.23  &  10.32 &  9.53 \\ \vspace{-0.1cm}
NGC 5033                         & Sy2    &  0.00292 &    12.36  &  8.84 & 10.24 \\ \vspace{-0.1cm}
NGC 4151                         & Sy2    &  0.00331 &    14.02  &  10.37 &  9.63 \\ \vspace{-0.1cm}
NGC 1365\tablenotemark{b}           & Sy2    &  0.00545 &    23.12  &  10.29 & 10.84 \\ \vspace{-0.1cm}
NGC 3079 			& Sy2    &  0.00372 &    15.76  & 10.87  & 10.64 \\ \vspace{-0.1cm}
NGC 1068                         & Sy2    &  0.00379 &    16.06  & 11.36  & 11.13 \\ \vspace{-0.1cm}
NGC 7582                         & Sy2    &  0.00525 &    22.28  & 11.17  & 10.62 \\ \vspace{-0.1cm}
NGC 7314                         & Sy2    &  0.00476 &    20.18  & \ldots  &  9.78 \\ \vspace{-0.1cm}
NGC 5506                         & Sy2    &  0.00618 &    26.24  & 10.71  & 10.21 \\ \vspace{-0.1cm}
NGC 4418                         & Sy2    &  0.00729 &    30.98  & 12.50  & 11.09 \\ \vspace{-0.1cm}
NGC 4593                         & Sy1    &  0.00900 &    38.30  &  \ldots & 10.25 \\ \vspace{-0.1cm}
NGC 3783                         & Sy1    &  0.00973 &    41.43  &  10.51 & 10.18 \\ \vspace{-0.1cm}
Mrk 3                            	& Sy2    &  0.01350 &    57.65  & 9.78  & 10.41 \\ \vspace{-0.1cm}
IC 4329 A                        	& Sy1    &  0.01605 &    68.68  & 10.22  & 10.35 \\ \vspace{-0.1cm}
NGC 7469                         & Sy1    &  0.01631 &    69.80  & 11.02  & 11.35 \\ \vspace{-0.1cm}
NGC 1275                         & Sy2    &  0.01756 &    75.23  & 10.85 & 10.89 \\ \vspace{-0.1cm}
IRAS F18325-5926           & Sy2    &  0.02002 &     85.93  & 10.51  & 10.70 \\ \vspace{-0.1cm}
Mrk 273                          & Sy2 &  0.03777 & 164.34  & $>12.13$  & 12.10 \\ \vspace{-0.1cm}
Mrk 231                             & Sy1 &  0.04217 & 184.09  & $>12.26$  & 12.34 \\  \hline \vspace{0.1cm} 
{\it Luminous Infrared Galaxies}   &    &   &   &  &  \\ \hline 
Arp 299                          & HII    &  0.01030 &    43.87  &  11.92 & 11.74 \\ \vspace{-0.1cm}
Arp 220                          & Lin    &  0.01813 &    77.70  & 12.53  & 12.13 \\ \vspace{-0.1cm}
NGC 6240                         & Lin    &  0.02448 &   105.44  & 11.28  & 11.73 \\ \vspace{-0.1cm}
UGC 5101                         & Lin    &  0.03937 &   171.51  & 11.32  & 11.91 \\ \vspace{0.1cm}
IRAS 13120-5453              & Sy2    &  0.03076 & 133.12  & 11.99 & 12.20 \\ \vspace{-0.1cm}
IRAS 20551-4250                 & HII &  0.04294 &   187.56  &  11.64 & 11.93 \\ \vspace{-0.1cm}
IRAS F05189-2524            & Sy2  &  0.04256 &   185.85  & 11.78  & 11.97 \\ \vspace{-0.1cm}
IRAS 17208-0014                  & HII &  0.04281 &  186.97  & 11.97 & 12.36 \\ \vspace{-0.1cm}
IRAS F10565+2448            & HII    &  0.04310 &  188.28  & 11.54  & 11.97 \\ \vspace{-0.1cm}
IRAS 15250+3609             & Lin    &  0.05515 &   243.10  & 11.60  & 11.94 \\ \vspace{-0.1cm}
IRAS F08572+3915           & Lin    &  0.05835 &   257.80  & 11.73  & 12.04 \\ \vspace{-0.1cm}
IRAS 09022-3615              & HII    &  0.05964 &   263.75  & 11.36 & 12.23 \\ \vspace{-0.1cm}
IRAS 23365+3604             & Lin    &  0.06448 &   286.17  &  11.65 & 12.09 \\ \vspace{-0.1cm}
IRAS 19542+1110            & \ldots &  0.06495 &   288.35  &  11.75 & 12.01 \\ \vspace{-0.1cm}
IRAS F14378-3651            & Lin    &  0.06764 &   300.88  & 11.81 & 12.11 \\ \vspace{-0.1cm}
IRAS F12112+0305          & Lin    &  0.07332 &   327.49  & 10.88  & 12.27 \\ \vspace{-0.1cm}
IRAS 22491-1808              & HII    &  0.07776 &   348.43  &  11.26 & 12.11 \\ \vspace{-0.1cm}
IRAS F14348-1447            & Lin    &  0.08300 &   373.30  & 10.48 & 12.29 \\ \vspace{-0.1cm}
IRAS F19297-0406            & HII    &  0.08573 &   386.32  & 11.67 & 12.35 \\ \vspace{-0.1cm}
IRAS 07251-0248              & \ldots &  0.08756 &  395.08  & 11.49  & 12.31 \\ 
IRAS 15206+3342 		& HII    &  0.12441 &   575.75  & \ldots & 12.07 \\
\enddata
\tablenotetext{a}{Distances were calculated using the redshift given in the NASA/IPAC Extragalactic Database (and references therein) assuming a flat $\Lambda$CDM cosmology with $\Omega_{\rm m} = 0.27$ and a Hubble constant $H_{0}=73$~km~$s^{-1}$~Mpc$^{-1}$.}
\tablenotetext{b}{These systems exhibit both an AGN (Sy 2) and a circumnuclear starburst.}
\end{deluxetable*}

\startlongtable
\begin{deluxetable*}{lcccc}\label{table:Observations summary}
\tablecaption{Observations summary \label{table:Observations Summary}}
\tablehead{
\colhead{Source} & \colhead{Position} & \colhead{R.A.} & \colhead{Dec.} & \colhead{{\it Herschel} OBSID} \\
\colhead{} & \colhead{} & \colhead{J2000} & \colhead{J2000} & \colhead{} \\
}
\vspace{-0.3cm}
\startdata  
M 82    & Central region   &  09 55 52.22  &  $+$69 40 46.9 & 1342186798--9                  \\ \vspace{-0.1cm}
            & Southern outflow &  09 55 56.00  &  $+$69 40 04.0 & 1342208945         \\ \vspace{-0.1cm}
            & Northern outflow &  09 55 48.80  &  $+$69 41 27.0 & 1342208946     \\ \vspace{-0.1cm}
M 83    & Central region           &  13 37 00.92  &  $-$29 51 56.7 & 1342203445--6        \\ \vspace{-0.1cm}
            & Bar--spiral arm  &  13 36 54.20  &  $-$29 53 01.5 & 1342203447--8     \\ \vspace{-0.1cm}
             & Eastern arm       &  13 37 07.20  &  $-$29 52 46.1 & 1342203449--50    \\ \vspace{-0.1cm}
NGC 253  &   &  00 47 33.12  &  $-$25 17 17.6 & 1342199414--5                         \\ \vspace{-0.1cm}
NGC 1808          &   &  05 07 42.34  &  $-$37 30 47.0 & 1342214356, 1342219440  \\ \vspace{-0.1cm}
NGC 3256          &   &  10 27 51.27  &  $-$43 54 13.8 & 1342210383--4                 \\ \vspace{-0.1cm}
Antennae         &  NGC4038 &  12 01 52.98  &  $-$18 52 09.9 & 1342199403--6      \\ \vspace{-0.1cm}
	      	       &  NGC4039 &  12 01 54.30  &  $-$18 53 02.0 & 1342210820--1     \\ \vspace{-0.1cm}
	      	       &  Overlap &  12 01 55.20  &  $-$18 52 29.0 & 1342210822--3     \\ \vspace{-0.1cm}
NGC 4945          &   &  13 05 27.48  &  $-$49 28 05.6 & 1342211694,1342212220      \\ \vspace{-0.1cm}
NGC 7552          &   &  23 16 10.77  &  $-$42 35 05.4 & 1342210399--400      \\ \vspace{-0.1cm}
NGC 1365          &   &  03 36 36.37  &  $-$36 08 25.4 & 1342191294--5      \\ \vspace{-0.1cm}
NGC 3738          &   &  11 39 01.72  &  $-$37 44 18.9 & 1342212225--6      \\ \vspace{-0.1cm}
NGC 4051          &                  &  12 03 09.61  &  $+$44 31 52.8 & 1342197812, 1342199142        \\ \vspace{-0.1cm}
NGC 4151          &                  &  12 10 32.58  &  $+$39 24 20.6 & 1342198305--6          \\ \vspace{-0.1cm}
NGC 4593          &                  &  12 39 39.42  &  $-$05 20 39.3 & 1342198302--3       \\ \vspace{-0.1cm}
NGC 5033          &                  &  13 13 27.53  &  $+$36 35 38.1 & 1342197821, 1342197908       \\ \vspace{-0.1cm}
NGC 5506          &                  &  14 13 14.87  &  $-$03 12 27.0 & 1342213758--9     \\ \vspace{-0.1cm}
NGC 7469          &                  &  23 03 15.62  &  $+$08 52 26.4 & 1342187847, 1342211171     \\ \vspace{-0.1cm}
IC 4329 A         &                  &  13 49 19.26  &  $-$30 18 34.0 & 1342203451, 1342223809    \\ \vspace{-0.1cm}
Cen A             &                  &  13 25 27.61  &  $-$43 01 08.8 & 1342202589--90     \\ \vspace{-0.1cm}
Circinus          &                  &  14 13 09.91  &  $-$65 20 20.5 & 1342191297--8      \\ \vspace{-0.1cm}
NGC 1068          &                  &  02 42 40.71  &  $-$00 00 47.8 & 1342191153--4       \\ \vspace{-0.1cm}
NGC 1275          &                  &  03 19 48.16  &  $+$41 30 42.1 & 1342191350--1     \\ \vspace{-0.1cm}
NGC 1386          &                  &  03 36 46.22  &  $-$35 59 57.3 & 1342192137--8    \\ \vspace{-0.1cm}
NGC 7314          &                  &  22 35 46.23  &  $-$26 03 00.9 & 1342197892, 1342210403       \\ \vspace{-0.1cm}
NGC 7582          &                  &  23 18 23.50  &  $-$42 22 14.0 & 1342210401, 1342211202      \\ \vspace{-0.1cm}
Mrk 3             &                  &  06 15 36.36  &  $+$71 02 15.1 & 1342220932, 1342219851, 1342220607    \\ \vspace{-0.1cm}
NGC 3079          &                  &  10 01 57.80  &  $+$55 40 47.2 & 1342221390--1      \\ \vspace{-0.1cm}
IRAS F18325-5926  &                  &  18 36 58.29  &  $-$59 24 08.6 & 1342192142--3   \\ \vspace{-0.1cm}
NGC 4418          &                  &  12 26 54.62  &  $-$00 52 39.2 & 1342187780--1, 1342210830     \\ \vspace{-0.1cm}
Arp 299           &                  &  11 28 32.20  &  $+$58 33 44.0 & 1342199419, 1342199421      \\ \vspace{-0.1cm}
Arp 220           &                  &  15 34 57.12  &  $+$23 30 11.5 & 1342191304--13, 1342212597 (1342238928--37)     \\ \vspace{-0.1cm}
Mrk 231           &                  &  12 56 14.23  &  $+$56 52 25.2 & 1342186811, 1342189280 (1342185482--502, 1342186037--53) \\ \vspace{-0.1cm}
Mrk 273           &                  &  13 44 42.11  &  $+$55 53 12.7 & 1342207801--2      \\ \vspace{-0.1cm}
NGC 6240          &                  &  16 52 58.89  &  $+$02 24 03.4 & 1342216622--3      \\ \vspace{-0.1cm}
UGC 5101          &                  &  09 35 51.65  &  $+$61 21 11.3 & 1342208948--9     \\ \vspace{-0.1cm}
IRAS 17208-0014   &                  &  17 23 21.96  &  $-$00 17 00.9 & 1342229692--3 (1342190690--711)   \\ \vspace{-0.1cm}
IRAS 20551-4250   &                  &  20 58 26.79  &  $-$42 39 00.3 & 1342208933--5   \\ \vspace{-0.1cm}
IRAS 23128-5919   &                  &  23 15 46.78  &  $-$59 03 15.6 & 1342210394--5      \\ \vspace{-0.1cm}
IRAS 23365+3604   &                  &  23 39 01.27  &  $+$36 21 08.7 & 1342212514--6    \\ \vspace{-0.1cm}
IRAS 15250+3609   &                  &  15 26 59.40  &  $+$35 58 37.5 & 1342213751--3  \\ \vspace{-0.1cm}
IRAS 19542+1110   &                  &  19 56 35.44  &  $+$11 19 02.6 & 1342208915--6   \\ \vspace{-0.1cm}
IRAS 22491-1808   &                  &  22 51 49.26  &  $-$17 52 23.4 & 1342211824--5       \\ \vspace{-0.1cm}
IRAS F05189-2524  &                  &  05 21 01.47  &  $-$25 21 45.4 & 1342219441--4      \\ \vspace{-0.1cm}
IRAS 07251-0248   &                  &  07 27 37.55  &  $-$02 54 54.1 & 1342207823, 1342207825--6     \\ \vspace{-0.1cm}
IRAS F08572+3915  &                  &  09 00 25.39  &  $+$39 03 54.4 & 1342208951--3, 1342208955--6   \\ \vspace{-0.1cm}
                  &                  &               &                & (1342184666--80, 1342184686--92, 1342186055--4)   \\ \vspace{-0.1cm}
IRAS 09022-3615   &                  &  09 04 12.70  &  $-$36 27 01.1 & 1342209402--5   \\ \vspace{-0.1cm}
IRAS F10565+2448  &                  &  10 59 18.14  &  $+$24 32 34.3 & 1342207787--9       \\ \vspace{-0.1cm}
IRAS F12112+0305  &                  &  12 13 46.00  &  $+$02 48 38.0 & 1342210831--3  \\ \vspace{-0.1cm}
IRAS 13120-5453   &                  &  13 15 06.35  &  $-$55 09 22.7 & 1342214628--9 (1342190689-710)   \\ \vspace{-0.1cm}
IRAS F14348-1447  &                  &  14 37 38.37  &  $-$15 00 22.8 & 1342203457, 1342224242--3     \\ \vspace{-0.1cm}
IRAS F14378-3651  &                  &  14 40 59.01  &  $-$37 04 32.0 & 1342204337--8      \\ \vspace{-0.1cm}
IRAS F19297-0406  &                  &  19 32 22.26  &  $-$04 00 00.7 & 1342208890--2     \\ \vspace{-0.1cm}
IRAS 15206+3342   &                  &  15 22 38.04  &  $+$33 31 35.9 & (1342189950--2)      \\ \\  
\enddata
\tablecomments{OBSIDs in parenthesis are from PV observations.}
\end{deluxetable*}

\begin{table*}
\caption{Ratio between the flux measured in the spaxel with the brightest FIR continuum and the total flux in the $5\times5$ spaxel array for SHINING galaxies with $D<50$~Mpc.}
\label{table:fractions}      
\centering          
\begin{tabular}{l c c c c c c c c c c c c | c}   
\hline\hline       
Source & \multicolumn{2}{c}{\niii} & \multicolumn{2}{c}{\oi} & \multicolumn{2}{c}{\oiii} & \multicolumn{2}{c}{\nii} & \multicolumn{2}{c}{\oii} & \multicolumn{2}{c}{\cii~158~$\mu$m} & \\ 
\cline{2-3} \cline{4-5} \cline{6-7} \cline{8-9} \cline{10-11} \cline{12-13}  
       & Line & Cont. & Line & Cont. & Line & Cont. & Line & Cont. & Line & Cont. & Line & Cont. & \\ 
\hline
Point source                 & 0.72   & 0.72   &  0.70   & 0.70   & 0.69   & 0.69   & 0.62   & 0.62   & 0.55   & 0.55    &  0.50    & 0.50  & \\  
50$\%$ point source          & 0.38   & 0.38    & 0.37   & 0.37   &  0.37   & 0.37   &   0.33   & 0.33   &   0.30   & 0.30    &  0.27    & 0.27 &  Extended (E) \\   
Ext.~uniform source          & 0.04   & 0.04   & 0.04   & 0.04   &   0.04   & 0.04   &   0.04   & 0.04   &   0.04   & 0.04    &   0.04    & 0.04  & or compact (C)? \\
\hline
{M 82 Central Reg.}            & 0.22   & 0.19   &  0.17   & 0.17   &  0.18   & 0.16   & 0.15   & 0.13   &  0.14   & 0.12      & 0.10    & 0.13  & E\\ \vspace{-0.1cm} 
{M 82 S. outflow}  & \ldots & \ldots &  \ldots & \ldots &  \ldots & \ldots &  0.17   & 0.14   &  \ldots & \ldots    & \ldots  & \ldots & E\\ \vspace{-0.1cm}
{M 82 N. outflow}  & \ldots & \ldots &  \ldots & \ldots &   \ldots & \ldots &  0.21   & 0.16   &   \ldots & \ldots    & \ldots  & \ldots & E\\ \vspace{-0.1cm}
{M 83 Central Reg.}            & 0.40   & 0.31   &  0.27   & 0.29   &   0.30   & 0.26   &   0.25   & 0.20   &   0.22   & 0.19      & 0.17    & 0.19  & E \\ \vspace{-0.1cm}
{M 83 Bar-spiral}   & 0.09   & 0.08   &  0.08   & 0.07     & \ldots & \ldots & 0.08   & 0.07   &   \ldots & \ldots    & 0.06    & 0.06  & E\\ \vspace{-0.1cm}
{M 83 East. arm}        & 0.37   & 0.17   &  0.17   & 0.18     & \ldots & \ldots &  0.17   & 0.10   &   \ldots & \ldots    & 0.11    & 0.09  & E\\ \vspace{-0.1cm}
NGC 253                      & 0.48   & 0.54   &  0.33   & 0.53   &   0.35   & 0.45   &  0.33   & 0.36   &   0.31   & 0.29    & 0.18    & 0.27  & C \\ \vspace{-0.1cm}
NGC 1808                     & 0.43   & 0.46   &  0.38   & 0.45   &  0.27   & 0.31   &   0.26   & 0.25   &   0.25   & 0.23      & 0.20    & 0.23  & C \\ \vspace{-0.1cm}
NGC 3256                     & 0.36   & 0.42   &  0.33   & 0.42   &   0.27   & 0.37   &  0.29   & 0.30   &   0.27   & 0.27      & 0.20    & 0.26  & C\\ \vspace{-0.1cm}
{Arp 299}                & 0.16   & 0.20   &  0.15   & 0.20   &   0.17   & 0.25   &  0.18   & 0.24   &   0.21   & 0.21      & 0.13    & 0.20  & E \\ \vspace{-0.1cm}
{NGC 4038}               & 0.12   & 0.14   &  0.10   & 0.15     & 0.04   & 0.10   &  0.13   & 0.09     & 0.09   & 0.09      & 0.07    & 0.11 &  E \\ \vspace{-0.1cm}
{NGC 4039}               & 0.23   & 0.25   &  0.15   & 0.24     & \ldots & \ldots &   0.12   & 0.15     & \ldots & \ldots    & 0.09    & 0.12  & E \\ \vspace{-0.1cm}
{Overlap}                & 0.11   & 0.14   &  0.15   & 0.14   &   \ldots & \ldots &   0.08   & 0.11   &   \ldots & \ldots  &   0.13    & 0.12  & E \\ \vspace{-0.1cm}
NGC 4945                     & 0.48   & 0.54   &  \ldots & 0.53   &   0.42   & 0.48   &   0.39   & 0.40     & 0.40   & 0.34      & 0.20    & 0.31 & C \\ \vspace{-0.1cm}
NGC 7552                     & 0.54   & 0.63   &  0.56   & 0.59   &   0.48   & 0.55   &   0.48   & 0.45     & 0.43   & 0.40      & 0.31    & 0.37 & C \\ \vspace{-0.1cm}
{NGC 1365}               & 0.27   & 0.24   &  0.22   & 0.24   &   0.29   & 0.23   &   0.21   & 0.21    & 0.21   & 0.20      & 0.17    & 0.20 & E \\ \vspace{-0.1cm}
NGC 3783                     & \ldots & 0.51   &  0.56   & 0.43   &   0.43   & 0.33   &   0.09   & 0.13     & 0.49   & 0.09      & 0.07    & 0.08 & C \\ \vspace{-0.1cm}
NGC 4051                     & \ldots & 0.40   &  0.47   & 0.64   &   0.38   & 0.64   &   0.44   & 0.29     & 0.42   & 0.24      & 0.21    & 0.20 & C \\ \vspace{-0.1cm}
NGC 4151                     & 0.54   & 0.65   &  0.74   & 0.56   &   0.57   & 0.49   &   0.47   & 0.25     & 0.49   & 0.21      & 0.29    & 0.17 & C \\ \vspace{-0.1cm}
NGC 4593                     & \ldots & \ldots &  0.63   & 0.59   &   0.46   & 0.57   &   0.37   & 0.35     & 0.41   & 0.30      & 0.31    & 0.29 & C \\ \vspace{-0.1cm}
{NGC 5033}               & 0.16   & 0.16   &  0.19   & 0.14     & 0.13   & 0.12   &  0.07   & 0.09   & 0.14   & 0.09      & 0.07    & 0.09 & E \\ \vspace{-0.1cm}
NGC 5506                     & 0.41   & 0.50   &  0.51   & 0.41     & 0.50   & 0.57   &   0.36   & 0.39    & 0.44   & 0.33    & 0.34    & 0.30 & C \\ \vspace{-0.1cm}
NGC 7469                     & 0.55   & 0.63   &  0.59   & 0.60     & 0.68   & 0.61   &   0.51   & 0.48    & 0.49   & 0.42     & 0.36    & 0.41 & C \\ \vspace{-0.1cm}
{Cen A}                  & 0.11   & 0.22   &  0.31   & 0.21     & 0.08   & 0.17   &  0.07   & 0.14     & 0.22   & 0.13      & 0.08    & 0.13 & E \\ \vspace{-0.1cm}
{Circinus}               & 0.28   & 0.41   &  0.33   & 0.39     & 0.22   & 0.32   &   0.16   & 0.24     & 0.27   & 0.20      & 0.13    & 0.19 & E \\ \vspace{-0.1cm}
{NGC 1068}               & 0.42   & 0.31   & 0.29   & 0.31   & 0.27   & 0.23   &  0.10   & 0.15     & 0.19   & 0.13      & 0.07    & 0.12 & E \\ \vspace{-0.1cm}
NGC 1386                     & \ldots & 0.58   & 0.63   & 0.64   &   0.79   & 0.49   &  0.35   & 0.31     & 0.36   & 0.26      & 0.21    & 0.22 &  C \\ \vspace{-0.1cm}
NGC 7314                     & \ldots & 0.35   & 0.34   & 0.30   &   0.47   & 0.29   &  0.19   & 0.14     & 0.40   & 0.11      & 0.08    & 0.08 & C \\ \vspace{-0.1cm}
NGC 7582                     & 0.58   & 0.68   & 0.58   & 0.66   &   0.53   & 0.54   &  0.54   & 0.43     & 0.43   & 0.38     & 0.32    & 0.35 & C \\ \vspace{-0.1cm}
Mrk 3                        & \ldots & 0.60   &  0.65   & 0.44   &   0.43   & 0.51   &  0.34   & 0.31   &   0.48   & 0.22    &   0.33    & 0.18 & C \\ \vspace{-0.1cm}
NGC 3079                     & \ldots & \ldots &  \ldots & 0.46  & \ldots & \ldots &  \ldots & 0.29   & \ldots & \ldots  &  0.18    & 0.25 & C \\ \\
\hline                     
\end{tabular}
\tablecomments{Empty spaces indicate either that the line and continuum were not observed, or that the combined spectrum was too noisy to properly estimate the ratio. Galaxies with more than $\sim70\%$ of their line and continuum emission coming from an extended component are classified as extended (E).}
\end{table*}

Each {\it Herschel} observation is identified in the {\it Herschel} data archive with a different OBSID. Table~\ref{table:Observations summary} gives the OBSIDs of the SHINING observations used in this paper, together with the central coordinates of the PACS spectrometer FOV. Some of the targeted lines were also observed during the {\it Herschel} PV phase in some of our sample galaxies. We combined those observations with the SHINING data to improve the significance of the detections and upper limits. While the majority of the SHINING galaxies have observations that cover the six main FIR lines in the $\sim55-160~\mu$m range, for about 20\% of the galaxies at least one line was not observed due to time limitations (in most cases the \niii\ line, see Table~\ref{table:Global measurements} for details). Additional lines \citep[e.g., OH  119, 79, 65~$\mu$m;][]{rhc_fischer10,rhc_gonzalez-alfonso17} were observed in a subset of the targets but are not discussed here.

\section{Data reduction}\label{data:reduction}
         
The data were reduced using the standard PACS data reduction and calibration pipeline included in HIPE v13.0 (Herschel Interactive Processing Environment\footnote{HIPE is a joint development by the Herschel Science Ground Segment Consortium, consisting of ESA, the NASA Herschel Science Center, and the HIFI, PACS and SPIRE consortia.}; \citealt{rhc_ott10}). A detailed description of the pipeline can be found in the Herschel Explanatory Supplement Vol. III.\footnote{\url{https://www.cosmos.esa.int/web/herschel/legacy-documentation-pacs}} The most important steps involve flagging the saturated frames and glitches, the subtraction of the chopper OFF position from the ON position, and the flatfielding of the 16 PACS spectral pixels in each of the 25 spatial pixels, where we took as a reference the common continuum range seen by all pixels. Each nod position was reduced separately. The signal from the two nods was combined after the generation of the rebinned spectra. For the final calibration we normalized the signal to the telescope background and recalibrated it with a reference telescope spectrum obtained from dedicated Neptune observations during the PV phase.

\subsection{Integrated measurements}

The effective PACS spectrometer point spread function (PSF) is the convolution of the propagated {\it Herschel} telescope PSF ($\sim$6\arcsec--11\arcsec\ at the PACS wavelength range) with the shape of the spectrometer spatial pixel (hereafter referred to as a ``spaxel'', and approximately a 9.4\arcsec$\times$9.4\arcsec\ square in extent). As a result of that, each spaxel collects only a fraction of the signal and the rest is distributed in the neighboring spaxels. For point sources this fraction is well characterized both theoretically and from dedicated PV observations \citep{rhc_poglitisch10}. It  decreases with wavelength from $\sim$70\%\ at 57\,$\mu$m to $\sim$50\%\ at 158\,$\mu$m (for details see Table~\ref{table:fractions}). These corrections are useful to calculate the integrated signal in galaxies that can be considered point sources for {\it Herschel}. 

In the SHINING sample, 25 galaxies are closer than 60~Mpc and can be considered extended for the PACS spectrometer spatial resolution ($9\arcsec \lesssim 2.5$\,kpc for $D_{L} \lesssim 60$\,Mpc). We study the spatial distribution of the emission in these galaxies in Section~\ref{subsec:extended sources}. For the rest of the SHINING galaxies (the remaining 27 with $D_{L} > 60$\,Mpc), the global emission can be considered a point source, .i.e., the line and continuum fluxes in the central spaxel are $\gtrsim50\%$ of the total emission measured in the PACS $5\times5$ spaxel array.

The nominal point source correction assumes that the source is correctly centered in the central pixel. This was unfortunately not the case in some of the SHINING observations, either as a result of telescope pointing errors of $\sim$2\arcsec--3\arcsec\ or because the FIR coordinates of the galaxy were not known with enough accuracy before the observations (e.g., for IRAS 19542+1110). In those cases we scaled the line flux in the central pixel to the ratio between the integrated continuum in the 25 spatial pixels (detected with reasonable signal to noise ratio in most galaxies) and the central pixel continuum: 

\begin{equation}
F_{\rm{line}(\rm{total})} = F_{\rm{line}(\rm{central})} \times S_{\rm{cont}
(\rm{total})}/S_{\rm{cont}(\rm{central})}
\end{equation}

\noindent These corrections are minimally affected by pointing errors or the source size for unresolved or nearly unresolved, but assume that the line and continuum emission distributions are similar.   

To calculate the total integrated line and continuum emission in the SHINING galaxies more extended than the PSF we summed the signal from the 25 spaxels. Note that even this is not exactly accurate since the flux in each spaxel should be multiplied by a factor to take into account the flux spread outside of the array due to the PSF.

\begin{figure*}
\centering
\includegraphics[scale=0.07]{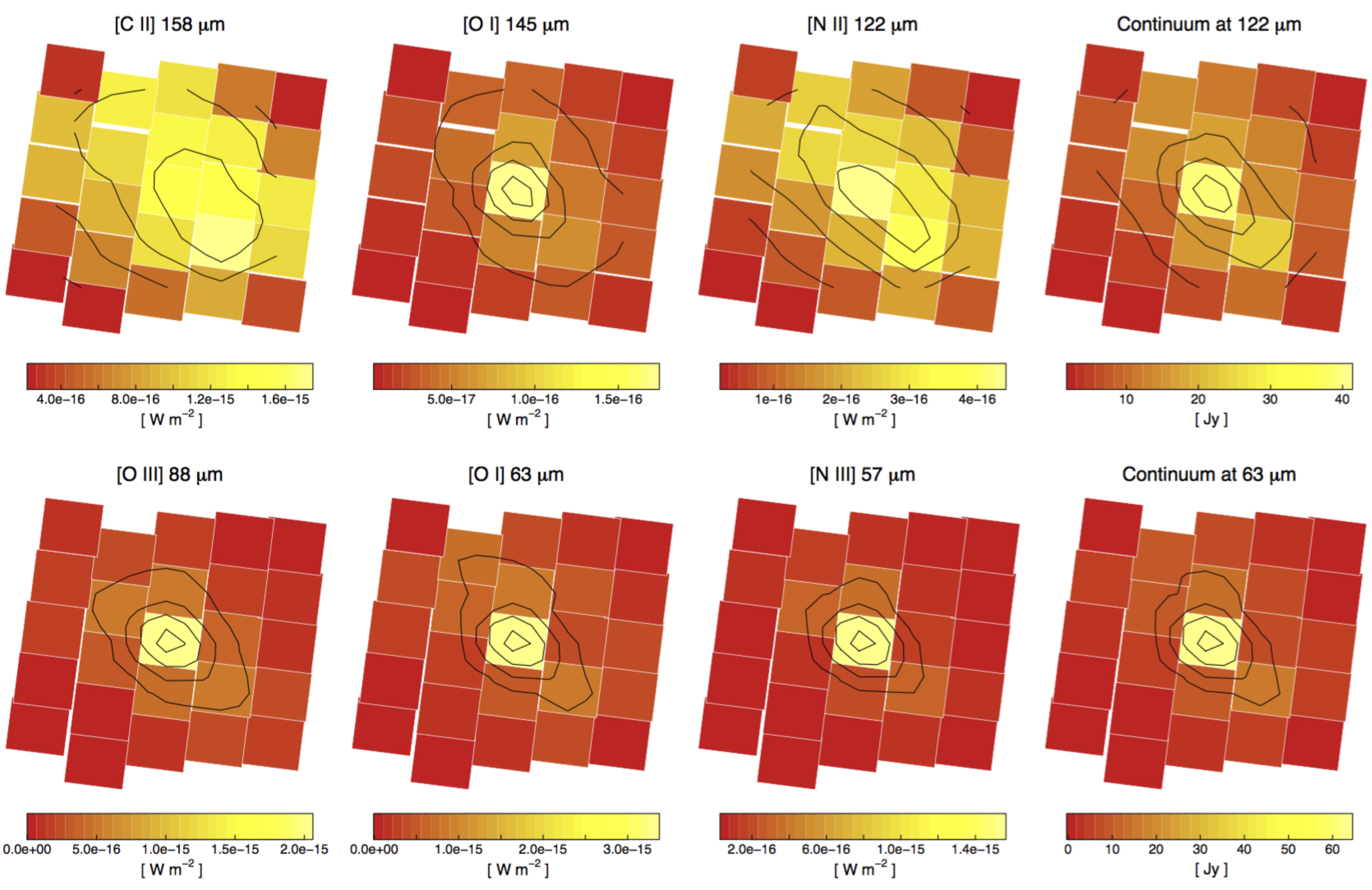}
\caption{Line and continuum maps of the Seyfert~2 galaxy NGC~1068. The maps are not corrected for the fact that the PACS spectrometer PSF increases with wavelength and is larger than the pixel size ($\sim$9.4\arcsec = 0.8\,kpc for $D_{L} = 18$\,Mpc; see sec.~\ref{subsec:extended sources} for details). Contours indicate regions where the signal is higher than 20\%, 40\%, 60\% and 80\% of the maximum value in the map.}
\label{Fig:NGC1068-maps} 
\end{figure*}

\subsection{Spatially resolved maps \label{subsec:extended sources}}

As we mentioned before, about 40\% of the SHINING sources are nearby and thus extended for the PACS spectrometer spatial resolution. This gives us the opportunity to study the distribution of the line and continuum emission in active and star forming galaxies on kiloparsec and sub-kiloparsec scales. 

As an example, Figure~\ref{Fig:NGC1068-maps} shows eight PACS maps of the prototypical Seyfert~2 galaxy NGC~1068 ($D_{L}=18$~Mpc, $9\arcsec \approx 0.8$~kpc). Around 30\% (15\%) of the total 63\,$\mu$m (122\,$\mu$m) continuum is detected in the central pixel, corresponding to the position of the AGN. We can also appreciate in Figure~\ref{Fig:NGC1068-maps} that the spatial distribution of the fine structure lines varies significantly among different lines. A large fraction ($\sim$30\%) of the \niii, \oiii, and \ois~63 and 145~$\mu$m emission is concentrated in the central pixel, revealing the effect of AGN radiation on the line excitation. In contrast, the \cii\ and \nii\ emission distributions are more extended and trace the parts of the disk where most of the recent star formation in the galaxy is taking place. In fact, the brightest \cii\ emission does not come from the galaxy center, but from a nearby pixel in the south-west direction. The position of this pixel coincides with a bright molecular gas clump seen in the CO interferometric maps of the galaxy \citep{rhc_planesas91,rhc_tacconi94,rhc_helfer95,rhc_schinnerer00}. This example shows that despite the relatively low spatial resolution of the PACS spectrometer and the reduced number of spatial pixels, it is still possible to extract valuable information from the SHINING maps. 

Given that observations of the SHINING galaxies were done in the pointed observing mode, the beam is undersampled (i.e., the beam is not sampled with at least the Nyquist resolution). This degrades the accuracy at which the morphology of observed sources can be reconstructed, and could introduce errors when comparing line and/or continuum maps at different wavelengths. This is the reason why for the analyses that involve spatially resolved regions we carefully select only galaxies for which the extent of the emission is significantly larger than one spaxel size. In these cases we expect that the flux that escapes a given spaxel is of the same order as the flux that enters that spaxel from nearby regions, which makes PSF effects less critical. Thus, the more extended a source is, the closer the fluxes detected in the pixels are to their intrinsic values.

From the group of 25 galaxies that can be considered extended for the PACS spectrometer, we identify those that can be seriously affected by the lack of correction for the PACS wavelength-dependent PSF. We do this by comparing the ratio between the emission measured in the spaxel with the brightest FIR continuum and the combined flux in the 25 spaxels to the expected peak-to-total ratios for three models of the emission distribution: a point source, an extended uniform source, and a combination of the two (50$\%$ point source, 50$\%$ extended)\footnote{In the case of the point source and mixed models the peak of emission is located in the central spaxel of the $5\times5$ spaxel array.}. Emission from extended uniform and half point, half extended sources have line and continuum central-to-total ratios in the $\sim0.04-0.4$ range. Inspection of Table~\ref{table:fractions} reveals that in 9 galaxies (4 starburst and 5 Seyfert) more than $\sim70\%$ of their total line and continuum emission can be attributed to an extended component. These are the galaxies we consider for our analysis of the spatially resolved emission (see the last column of Table~\ref{table:fractions} and Section~\ref{sec:resolved information}). The line fluxes for these spatially-resolved regions at the positions of each spaxel are reported in table \ref{table:Resolved measurements}.

\begin{figure*}
\plotone{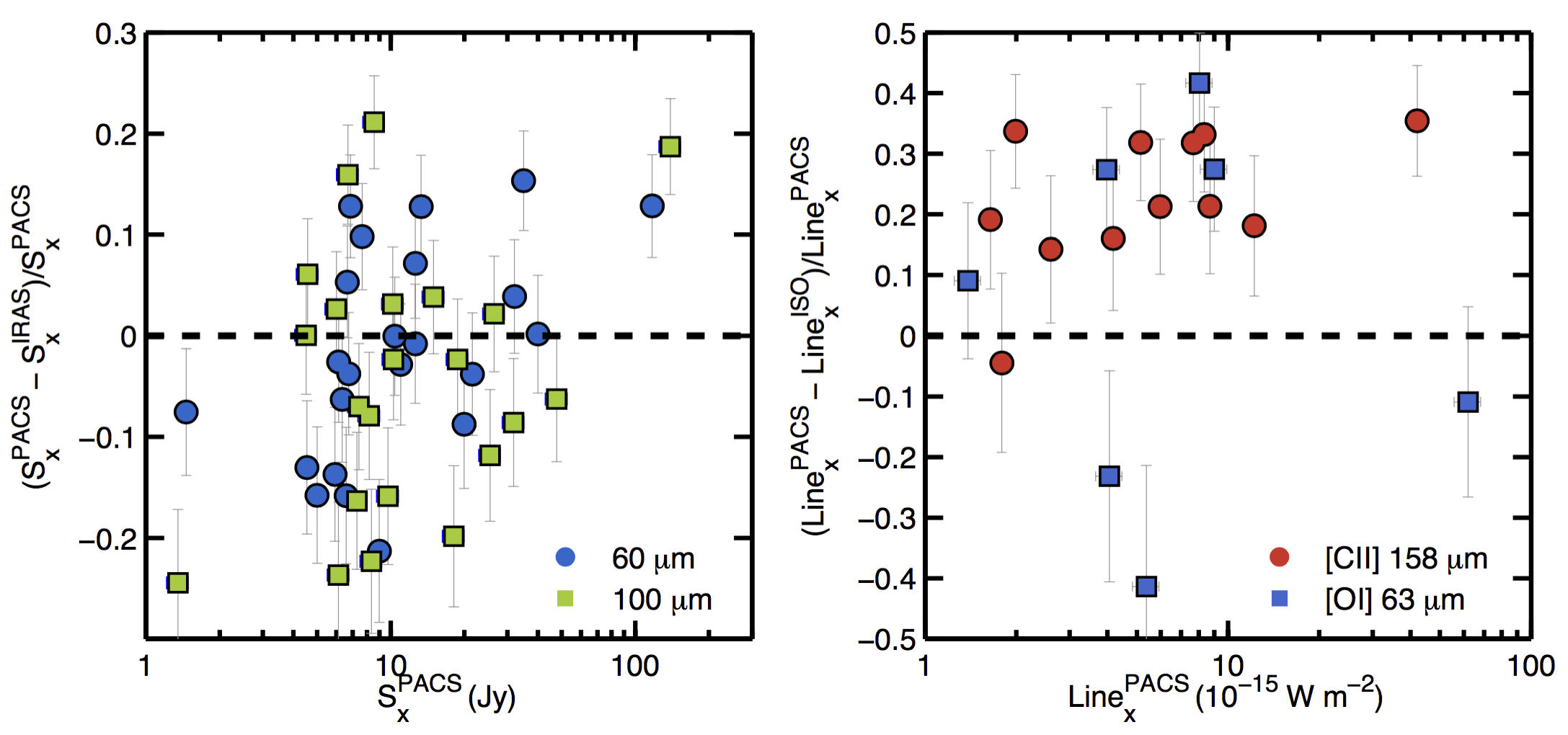} 
\caption{({\it Left}) Comparison between the continuum flux densities at 60~$\mu$m and 100~$\mu$m measured with IRAS and those estimated by us from the interpolation of the PACS continuum measurements. ({\it Right}) Comparison between the \cii\ and \oi\ line fluxes detected with ISO \citep{rhc_luhman03,rhc_brauher08} and those measured with the PACS spectrometer. \label{Fig:line comparison}}
\end{figure*}

The remaining 16 galaxies in Table~\ref{table:fractions} have more compact emission. In the cases of NGC~253 and NGC~4945, for example, more than half of the flux comes from the central 9.4\arcsec ($\sim$150\,pc) region. Their line emission distributions are also compact, with the exception of the \cii\ line, which is considerably more extended than the continuum at 158\,$\mu$m. The same is true for 10 of the 17 Seyfert galaxies. In this case their compactness is a combination of their higher luminosity distances compared to the \hii\ galaxies in the sample (see Table~\ref{table:SHINING sample}), and on average more compact coexisting star formation \citep{rhc_lutz17}. 

\section{Results}\label{sec:results}
          
Table~\ref{table:Global measurements} gives the integrated line fluxes for all the sources in the SHINING galaxy sample. We have detected most of the targeted lines. This represents a significant improvement with respect to previous observations of these lines in extragalactic sources by the Infrared Space Observatory (ISO) \citep{rhc_malhotra01,rhc_negishi01,rhc_luhman03,rhc_brauher08}. In those studies only the closest and brightest sources were observed and detected in the 6 fine structure lines \citep[][]{rhc_brauher08}. 

Errors in Table~\ref{table:Global measurements} are 1$\sigma$ uncertainties in the Gaussian line fits and exclude errors associated with the PACS absolute flux calibration. From the PACS manual, these numbers are 11\% ($50-70~\mu$m) and 12\% ($70-220~\mu$m), r.m.s. and 30\% peak to peak. In addition to the line fluxes, Table~\ref{table:Global measurements} gives the rest-frame \sblue\ and \sred\ continuum flux densities close to the \oi\ and \nii\ lines. Their ratio is a measure of the shape of the FIR spectral energy distribution of the source, which in turn depends on the average dust temperature and continuum optical depth in the galaxy. This ratio will be used in the following sections.

\subsection{Comparison with previous observations \label{subsec:comparison}}

In this section we compare our line and continuum measurements with previous Infrared Astronomical Satellite (IRAS) and ISO observations. To avoid the uncertainties associated with the different instrument FOVs, we have considered only ULIRGs in this comparison, which are approximately point sources for all instruments. 
   
In Fig.~\ref{Fig:line comparison} (left) we compare the IRAS~$S_{60\,\mu\mathrm{m}}$ and \sgreen\ flux densities, tabulated in the Revised Bright Galaxy Sample \citep{rhc_sanders03}, with the PACS estimated values from the interpolation of the continuum close to the fine structure lines and the parallel data. Differences between these two measurements are lower than $\sim25\%$ in most cases over two orders of magnitude in $S_{60\,\mu\mathrm{m}}$ and  $S_{100\,\mu\mathrm{m}}$.  

A similar comparison with previous ISO fine structure line observations \citep{rhc_luhman03,rhc_brauher08} is shown in Fig.~\ref{Fig:line comparison} (right). Differences can be as high as 40\% in this case. The PACS \cii\ line fluxes are on average $\sim$25\% higher than the ISO values, which we attribute to systematics in the absolute flux calibration between the two instruments at that wavelength. The larger scatter in the \oi\ line fluxes compared to the  $S_{60\,\mu\mathrm{m}}$ flux densities in Fig.~\ref{Fig:line comparison} is probably due to the ISO lower signal to noise detections.

We conclude that the typical uncertainty in the PACS absolute flux calibration is probably lower than 20\%. The trends discussed in this paper involve variations in the line and continuum measurements larger than a factor of $\sim2$, so they will not be strongly affected by these differences. 

\subsection{Global measurements \label{sec:global measurements}}

In order to study how the relative intensities of the fine structure lines depend on the global properties of galaxies, we begin by investigating the dependence of the ratio of the line fluxes to the total FIR fluxes on various galaxy properties. Here we adopt the FIR flux definition given in \citet{rhc_helou88}:

\begin{equation}\label{eq:fir}
\begin{split}
F_{\rm FIR}(42.5\,\mu\rm{m}-122.5\,\mu\rm{m}) = 1.26 \times 10^{-14} \\ 
\times(2.58\ S_{60\,\mu\rm{m}} + S_{100\,\mu\mathrm{m}}),
\end{split}
\end{equation}

\noindent where $F_{\rm FIR}$ is expressed in units of W\,m$^{-2}$, and $S_{60\,\mu\mathrm{m}}$ and $S_{100\,\mu\mathrm{m}}$ are in Jansky. We decided to continue using this version of the FIR flux to remain consistent with prior analyses, such as \cite{rhc_malhotra01} and \cite{rhc_diaz-santos13}. Note, however, that the FIR flux in the $42.5-122.5\,\mu\rm{m}$ range represents only about half of the total infrared emission emitted by the galaxy. Compared to the total FIR flux (or TIR) calculated in the 8 to 1000~$\mu$m wavelength range, $F_{\rm FIR}(42.5-122.5~\mu{\rm m})\approx(1/1.75) \times F_{\rm TIR}(8-1000~\mu{\rm m})$ \citep[see Apendix~E in][]{rhc_rhc15, rhc_vu12,rhc_delooze14}. 

\begin{figure*}
\includegraphics[scale=0.23]{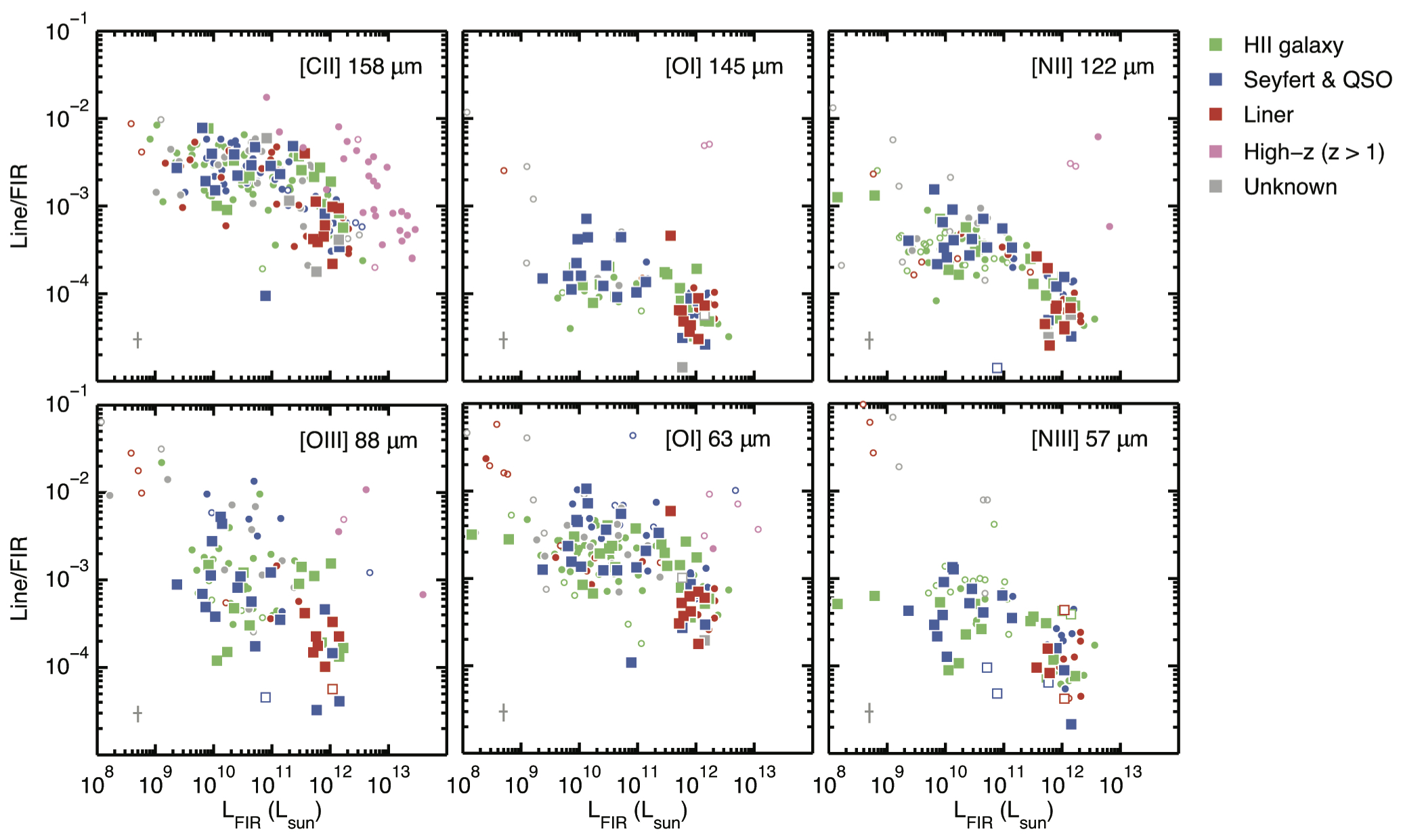}
\caption{Global line to FIR continuum ratios for galaxies with different optical activity classifications and high-$z$ ($z>1$) as a function of their far-infrared luminosity. Galaxies in the SHINING sample are marked with square symbols. Open symbols indicate 3-$\sigma$ upper limits to the line flux. Galaxies with $L_{\rm FIR}\gtrsim10^{11}$~L$_{\odot}$, irrespective of their types, tend to show low line to FIR continuum ratios. Typical errorbars are plotted in the lower-left corner.}
\label{Fig:deficits-Lfir}
\end{figure*}

In Fig.~\ref{Fig:deficits-Lfir} we plot the global line to FIR ratios measured in the SHINING galaxy sample as a function of the far-infrared luminosity. We also include in the figure a compilation of previous ISO extragalactic observations taken from the literature \citep{rhc_malhotra01,rhc_negishi01,rhc_luhman03,rhc_lutz03,rhc_brauher08}, and PACS \cii\ observations of local starburst, (U)LIRGs and AGN by \cite{rhc_sargsyan12} and \cite{rhc_farrah13}. High redshift observations ($z>1$) obtained with space and ground based facilities are also included \citep{rhc_dale04,rhc_iono06,rhc_maiolino09,rhc_walter09,rhc_wagg10,rhc_hailey-dunsheath10,rhc_ivison10,rhc_sturm10,rhc_stacey10,rhc_ferkinhoff10,rhc_ferkinhoff11,rhc_valtchanov11,rhc_deBreuck11,rhc_cox11,rhc_wagg12,rhc_walter12,rhc_riechers13,rhc_riechers14,rhc_debreuck14,rhc_capak15,rhc_oteo16,rhc_diaz-santos16}. The compilation of line and FIR continuum fluxes used in Figure~\ref{Fig:deficits-Lfir} is available via email request to the first author.

Galaxies in Fig.~\ref{Fig:deficits-Lfir} are color coded according to their redshift and optical nuclear activity classification following the Baldwin, Phillips and Terlevich (BPT) diagnostic diagram introduced in \cite{rhc_baldwin81} and \cite{rhc_veilleux87}. This figure is an updated version of Fig.~1 in \citet{rhc_gracia-carpio11}, now with PACS observations for the whole SHINING galaxy sample. We excluded galaxies that are spatially extended in the far-infrared at the IRAS spatial resolution since the ISO line observations were restricted to the galaxy centers and their line to FIR  ratios can only be considered lower limits \citep{rhc_helou88,rhc_sanders03,rhc_surace04}. Our SHINING observations are not affected by this problem because we have a direct estimate of the FIR continuum within the PACS FOV.

The line to FIR continuum ratios in galaxies with moderate far-infrared luminosities ($L_{\mathrm{FIR}} \lesssim 10^{11}\,L_{\odot}$) do not seem to depend strongly on $L_{\mathrm{FIR}}$. In the case of the \cii\ line, for example, the Kendall $\tau$ coefficient\footnote{The Kendall $\tau$ coefficient ($-1\leq \tau \leq1$) is a non-parametric test that indicates the degree of positive or negative correlation between two quantities.  If $\tau=1$, there is a full correlation, if $\tau=-1$, there is full anti-correlation, and if $\tau=0$, the two data sets are independent. In our analysis, together with the $\tau$ coefficient we list its associated p-value, which is the probability of getting a correlation as large as the observed value by random chance.} is $\tau=-0.02$ ($p=0.69$) for the subgroup of galaxies with $L_{\mathrm{FIR}} \lesssim 10^{11}\,L_{\odot}$. In general the dispersion is high and the ratios can vary by more than an order of magnitude, especially in the case of the high ionization lines.

Above $L_{\mathrm{FIR}} \approx10^{11}\,L_{\odot}$ we find an increasing number of galaxies with low line to FIR continuum ratios (the median line-to-FIR ratios of galaxies below $L_{\mathrm{FIR}}=10^{11}\,L_{\odot}$ are typically a factor of $\sim3$ higher than those of systems brighter than $L_{\mathrm{FIR}}=10^{11}\,L_{\odot}$). These line deficits were first identified with ISO in the \cii\ line \citep{rhc_malhotra97,rhc_luhman98}. Fig.~\ref{Fig:deficits-Lfir} shows that they affect all the PDR and \hii\ lines we have observed with PACS. In systems with $L_{\mathrm{FIR}} \gtrsim 10^{12}\,L_{\odot}$, and independent of their galaxy type, the relative intensity of the fine structure lines can drop by more than an order of magnitude compared to the average values found in galaxies with lower far-infrared luminosities. 

The FIR luminosity is, however, a bad proxy for the relative brightness of the fine structure lines. This is particularly true in the high redshift Universe, where galaxies with $L_{\mathrm{FIR}} \gtrsim 10^{12}\,L_{\odot}$ can have line to continuum ratios similar to those of local galaxies with $L_{\mathrm{FIR}} \lesssim 10^{11}\,L_{\odot}$ \citep[e.g.,][]{rhc_hailey-dunsheath10,rhc_stacey10,rhc_brisbin15}. As discussed in \citet{rhc_gracia-carpio11}, this can be understood in terms of the redshift evolution of the gas content in galaxies. Recent molecular gas observations of massive star forming galaxies at $z \sim 1$--2 have shown that their molecular gas fractions are typically three to ten times higher than in today's massive spirals \citep{rhc_daddi08,rhc_tacconi10,rhc_genzel15,rhc_tacconi17}. This explains part of the redshift evolution of the SFR--$M_{*}$ main sequence and why galaxies with otherwise similar ISM properties are more luminous at high redshift: they form more stars simply because they have more gas to form them. 

There is growing evidence that better proxies for the relative brightness of the fine structure lines are ${\rm SFE}=L_{\mathrm{FIR}}/M_{\mathrm{mol}}$ --which in star formation dominated galaxies is referred to as the molecular gas star formation efficiency (SFE)--, and the FIR surface brightness, $\Sigma_{\rm FIR}$  \citep{rhc_gracia-carpio11,rhc_diaz-santos14,rhc_lutz16,rhc_barcos17}. Both of these quantities are proportional to the strength of the UV field, most likely the main parameter controlling the efficiency of the conversion of UV radiation into gas heating. Motivated by these results, in the next sections we study the ratio between the FIR fine-structure lines and the FIR dust emission as a function of  $\Sigma_{\rm FIR}$ and \firmol\ in addition to $L_{\rm FIR}$.

\subsection{\cii~158~$\mu$m -- FIR ratio}

\begin{figure*}
\begin{center}
\includegraphics[scale=0.23]{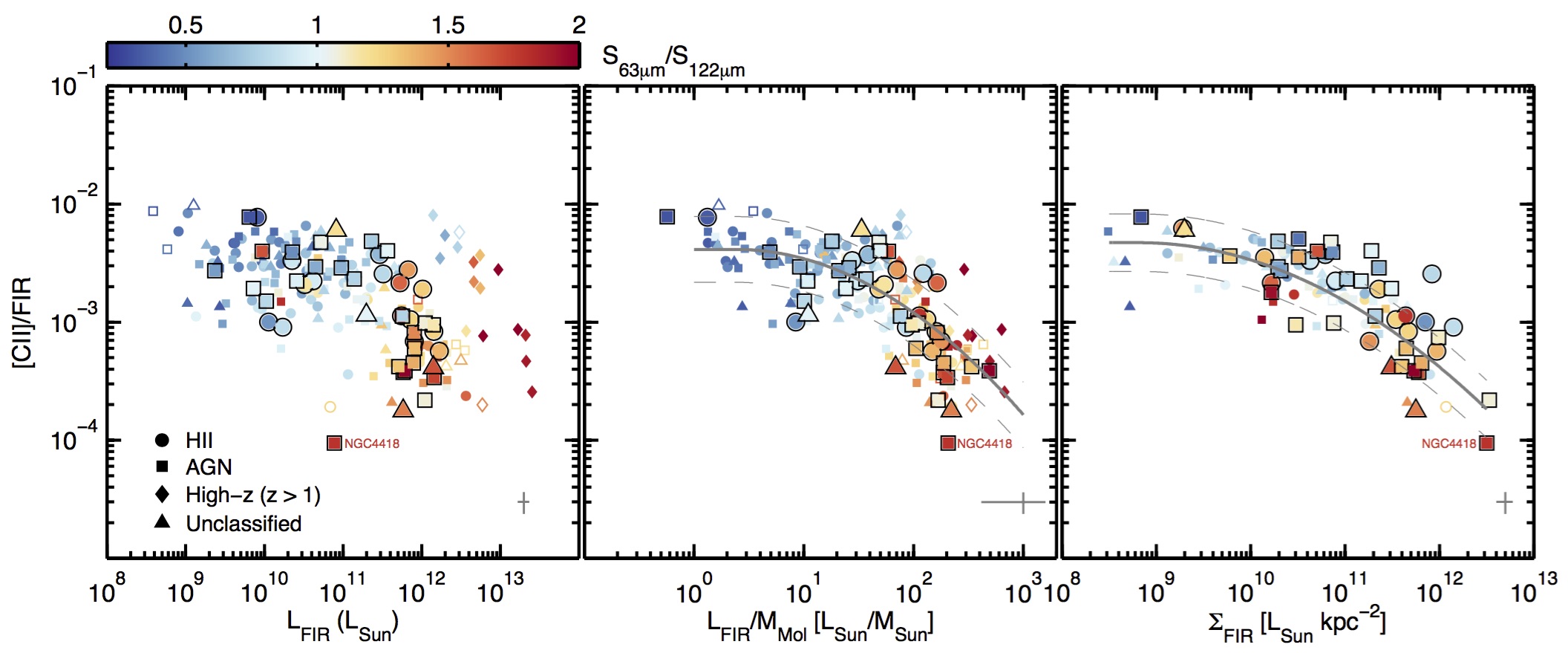}
\caption{Global \cii~158~$\mu$m line to FIR continuum ratio for different types of galaxies and high-$z$ ($z>1$) systems as a function of far-infrared luminosity (left), the ratio between far infrared luminosity and molecular gas mass (center), and infrared luminosity surface brightness (right). SHINING galaxies are marked with a black border. Open symbols indicate 3-$\sigma$ upper limits to the line flux and the color scale indicates the \sratio\ IR color. The gray lines in the second and third panels show a simple quadratic fit to the data and associated $\pm1$-$\sigma$ dispersion. The best fit parameters are listed in Table~\ref{tab:scaling}. Similar to \cite{rhc_diaz-santos17}, we set the fits to the maximum value of the quadratic equation below the value of \firmol\ or \sfir\ at which the maximum is reached. Typical errorbars are plotted in the lower-right corner.}\label{Fig:cii-deficits}
\end{center}
\end{figure*}

Figure~\ref{Fig:cii-deficits} shows the \cii\ to FIR ratio as a function of $L_{\rm FIR}$, $L_{\rm FIR}/M_{\rm mol}$,  and $\Sigma_{\rm FIR}$. The color scale indicates the \sratio\ IR color, to first order a proxy for the average dust temperature of a galaxy. As we already show in Figure~\ref{Fig:deficits-Lfir}, galaxies with low FIR luminosities have \cii/FIR ratios in the $\sim10^{-2}-10^{-3}$ range irrespective of their galaxy type. At FIR luminosities higher than $L_{\rm FIR}\gtrsim10^{11}$~$L_{\odot}$ we start to observe a drop in the \cii/FIR ratio, although accompanied by an increase in the scatter of a factor of $\sim2$. We also observe that, as first reported by \cite{rhc_malhotra97}, galaxies that exhibit a \cii-deficit also tend to have warm IR colors. High-$z$ galaxies included in this study seem to follow the same trends described before, but shifted towards higher FIR luminosities, i.e., these systems have $L_{\rm FIR}\gtrsim10^{12}~L_{\odot}$, but only those with $L_{\rm FIR}\gtrsim10^{13}~L_{\odot}$ show a \cii-deficit. This is consistent with the interpretation by \cite{rhc_stacey10} that luminous star-forming galaxies at high-$z$ seem to be scaled-up versions of local starbursts.  

The middle panel in Figure~\ref{Fig:cii-deficits} shows that galaxies with $L_{\rm FIR}/M_{\rm mol}\gtrsim80~L_{\odot}~M_{\odot}^{-1}$ tend to have lower \cii\ to continuum ratios than galaxies with more moderate $L_{\rm FIR}/M_{\rm mol}$ values \citep[see also][]{rhc_gracia-carpio11}. These galaxies also have warm IR colors, which is expected given the well known correlation between $L_{\rm FIR}/M_{\rm mol}$ and $S_{60~\mu \rm m}/S_{100~\mu \rm m}$ \citep[e.g.,][]{rhc_young89}: for star formation dominated galaxies, if the number of new formed stars per unit of molecular gas mass increases, the UV radiation heating the dust in the cloud increases, and the dust temperature rises. In addition, for a given $L_{\rm FIR}/M_{\rm mol}$, local and high redshift galaxies have similar \cii\ to FIR ratios. This suggests that any possible variation in the properties of the ISM with redshift does not significantly affect the excitation and emission of the \cii\ line. 

The last panel in Figure~\ref{Fig:cii-deficits} shows the correlation between the \cii/FIR ratio and $\Sigma_{\rm FIR}$. FIR sizes for the SHINING sample and about $\sim50$\% of the additional galaxies included in this study were measured by \cite{rhc_lutz16}. Despite the smaller number statistics, the correlation of \cii/FIR with \sfir\ \citep[1-$\sigma$ dispersion of dispersion 0.24~dex; see also][]{rhc_lutz16} is tighter than with $L_{\rm FIR}$ (1-$\sigma$ dispersion of 0.34~dex). One example to illustrate this point is the case of NGC~4418 (see labels in Figure~\ref{Fig:cii-deficits}). With a FIR luminosity of $\sim8\times10^{10}~{\rm L}_{\odot}$ and a \cii\ to FIR ratio of $\sim10^{-4}$ this galaxy is an outlier in the $L_{\rm FIR}-\cii/{\rm FIR}$ correlation, but its strong \cii\ deficit is consistent with its high IR surface brightness ($\Sigma_{\rm FIR}\approx3\times10^{12}~{\rm L}_{\odot}~{\rm kpc}^{-2}$). If we think of the FIR emission in star formation dominated galaxies as a proxy for star formation activity \citep[e.g.,][]{rhc_kennicutt12}, then the observed tight correlation between \cii/FIR and \sfir\ is consistent with other studies of nearby, star-forming galaxies that find a tighter relation between $\Sigma_{\rm [CII]}-\Sigma_{\rm SFR}$ than $L_{\rm [CII]}-{\rm SFR}$ \citep[e.g.,][]{rhc_rhc15}. 

In Paper~II \citep{rhc_rhc18b} we present a {\it toy model} we use to explore the connection between the radiation field intensity, the FIR surface brightness, and the strength of the PDR line emission in star-forming galaxies. We find that as galaxies become more compact and luminous, the \cii\ line emission becomes nearly independent of the radiation field strength, as opposed to the FIR intensity. In addition, the photoelectric heating efficiency decreases, and the ionization parameter reaches a limit value where the fraction of UV photons absorbed by dust in the \hii\ region becomes important.

\begin{figure}
\begin{center}
\includegraphics[scale=0.13]{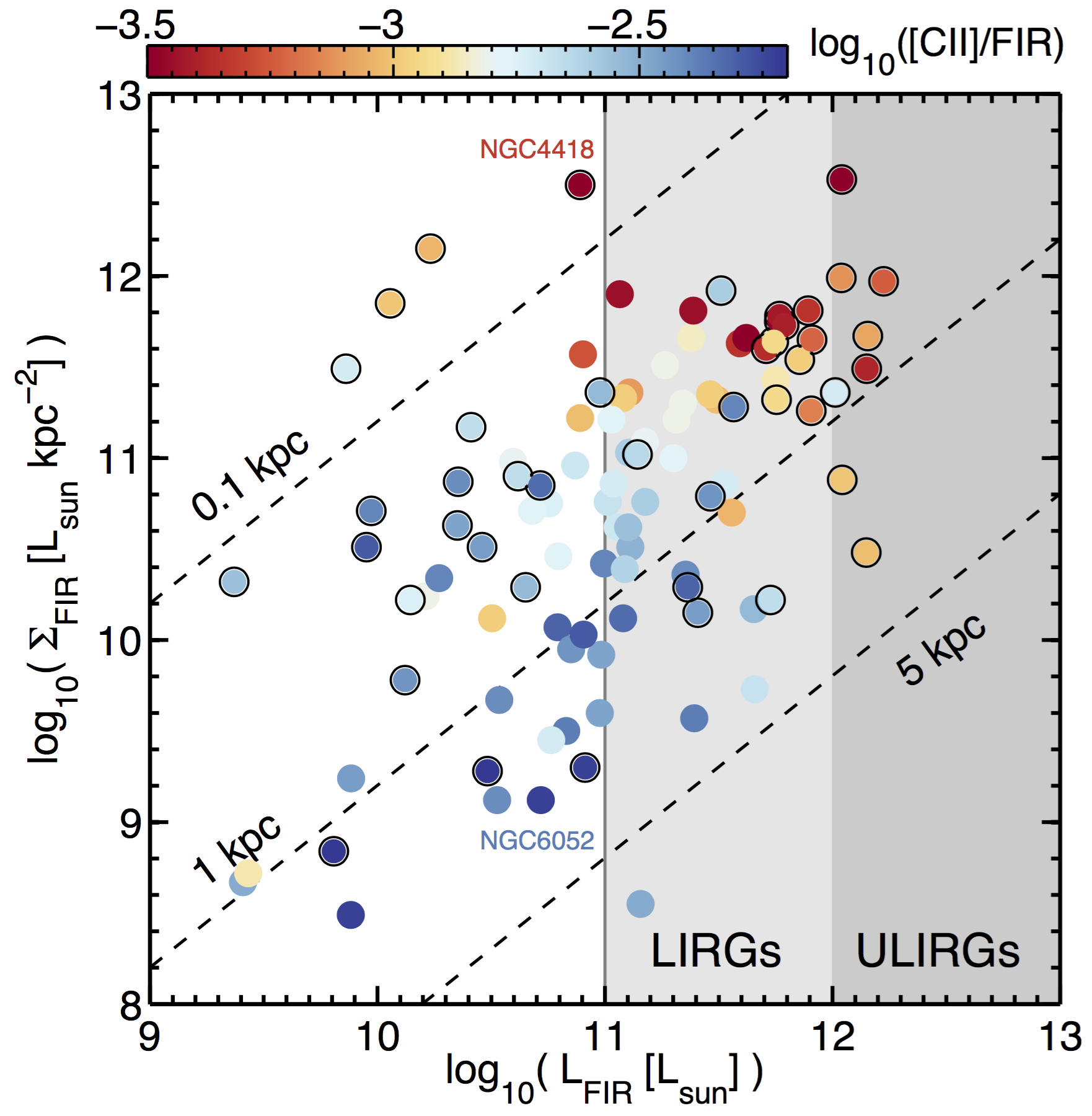}
\caption{FIR surface brightness ($\Sigma_{\rm FIR}$) as a function of FIR luminosity ($L_{\rm FIR}$) for individual galaxies color coded according to their \cii/FIR ratio. Diagonal gray lines show loci of constant IR radii, from 0.1~kpc to 5~kpc. SHINING galaxies are marked with a black border. The gray areas mark the space where galaxies are classified as LIRGs ($L_{\rm FIR}\geq10^{11}$~$L_{\odot}$) and ULIRGs ($L_{\rm FIR}\geq10^{12}$~$L_{\odot}$). It is wrong to assume that all LIRGs exhibit a \cii\ deficit: at a fixed luminosity there is a clear trend of decreasing \cii/FIR ratio with decreasing IR size of the galaxy. Errorbars are omitted as they are comparable to the size of the symbols.}\label{LvsS}
\end{center}
\end{figure}

The gray lines in the second and third panels of Figure~\ref{Fig:cii-deficits} show the best fit to the data. Similar to \cite{rhc_lutz16} and \cite{rhc_diaz-santos17}, we fit the data using a second order polynomial function. This function provides a better fit to our data than an exponential function. Note that we force the maxima of the best fitting quadratic functions to be equal to or less than the minimum of the abscissa values. The choice of using quadratic functions is motivated by the fact that they do a good job characterizing the shape of the empirical relations, and not because they analytically describe the physical processes behind these relations.

The best quadratic fit parameters and the maximum line-to-FIR ratio of the distribution are listed Table~\ref{tab:scaling}.

\begin{figure*}
\begin{center}
\includegraphics[scale=0.28]{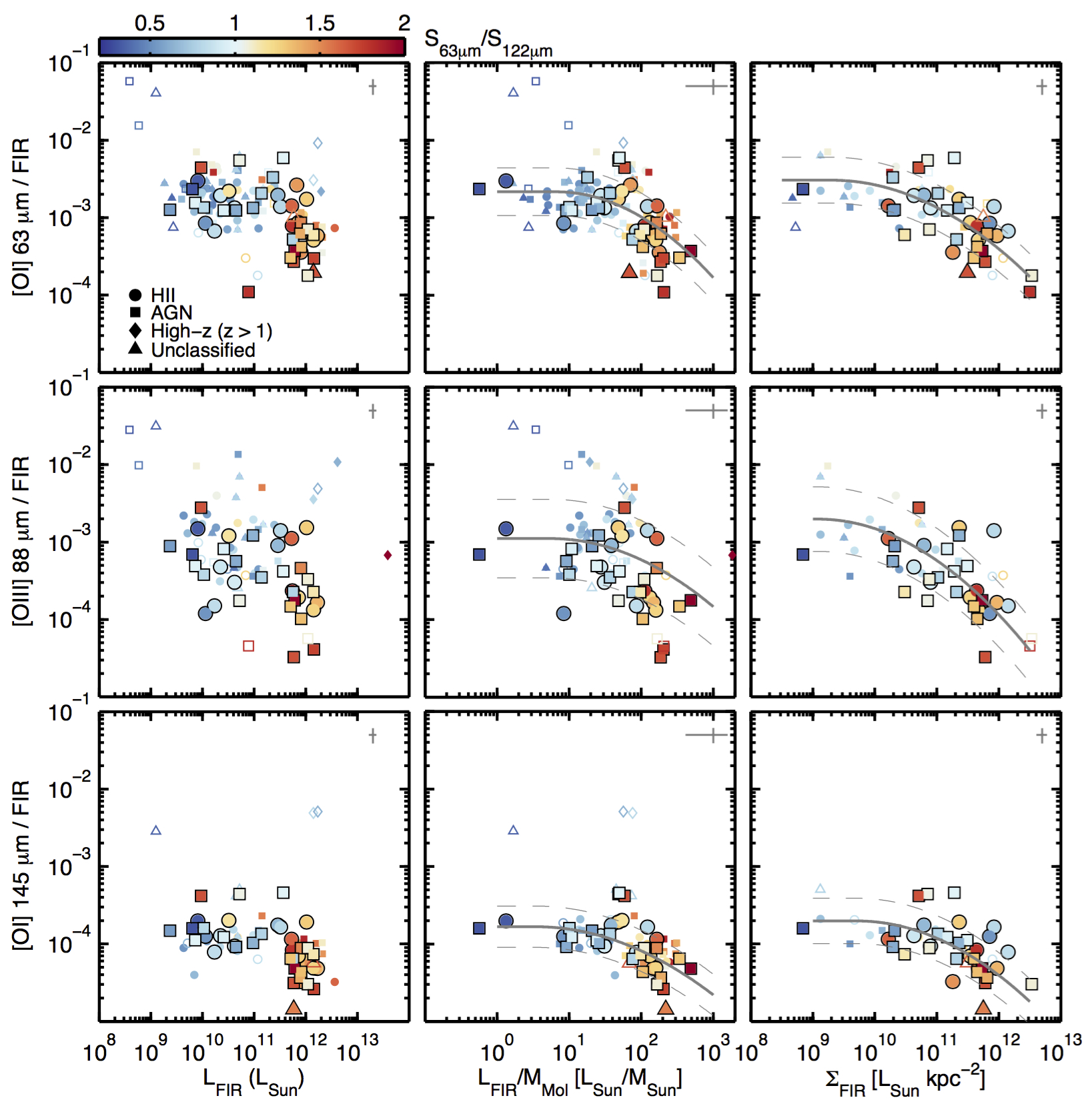}
\caption{Global \oi\ (top),  \oiii\ (middle), and \oii\ (bottom) line to FIR continuum ratio for different types of galaxies and high-$z$ ($z>1$) systems as a function of far-infrared luminosity (left), the ratio between far infrared luminosity and molecular gas mass (center), and infrared luminosity surface brightness (right). SHINING observations are marked with a black border. Open symbols indicate 3-$\sigma$ upper limits to the line flux and the color scale indicates the \sratio\ IR color. The gray lines show a simple quadratic fit to the data. The best fit parameters are listed in Table~\ref{tab:scaling}. We set the fits to the maximum value of the quadratic equation below the value of \firmol\ or \sfir\ at which the maximum is reached. Typical errorbars are plotted in the upper-right corner.}\label{Fig:ox-deficits}
\end{center}
\end{figure*}

An alternative representation of the first and last panel in Figure~\ref{Fig:cii-deficits} is shown in Figure~\ref{LvsS}. Here we plot the FIR luminosity versus the FIR surface brightness of individual galaxies. The color scale indicates the [CII] to FIR ratio, and the diagonal lines show loci of constant IR radii ($R_{e,70}=0.1$, 1, and 5~kpc). At a fixed luminosity we observe a trend of decreasing [CII]/FIR ratio as galaxies become more compact. To illustrate this point we compare NGC~6052 and NGC~4418. Both galaxies have similar FIR luminosities ($L_{\rm FIR}\sim5\times10^{10}$~$L_{\odot}$), but NGC~6052, with a FIR radius $\sim50$ times larger than NGC~4418 (2.4~kpc versus 0.05~kpc), has a [CII]/FIR ratio a factor of $\sim10^2$ higher.

\subsection{Oxygen lines to FIR ratio}

The two fine-structure lines of neutral oxygen, \oi\ and \oii, arise from neutral gas in PDRs and the warm neutral medium. Compared to \cii, these lines have higher upper level energies and critical densities and in warm, dense environments, they dominate the ISM cooling. The \oiii\ transition, on the other hand, only arises from \hii\ regions where photons with energies higher than $\sim35$~eV create O$^{++}$.

As shown for \cii\ in Figure~\ref{Fig:cii-deficits}, Figure~\ref{Fig:ox-deficits} shows the fine-structure oxygen lines to FIR ratios as a function of \lfir, \firmol, and \sfir. The color scale indicates the \sratio\ color, and different markers are used to differentiate galaxies according to their optical nuclear activity classification or redshift. AGN galaxies with high \ois/FIR and \oiii/FIR ratios generally have warm \sratio\ colors. (Seyfert galaxies with \oi, \oii, and \oiii\ line-to-FIR ratios below $10^{-3}$, $7\times10^{-5}$, and $3\times10^{-4}$, respectively, have warmer mean \sratio\ colors by factors of 1.5, 1.4, and 1.5 than the rest of the Seyfert systems). This could be interpreted as evidence for significant AGN contribution to the total FIR emission in the galaxy. There are however several \hii\ galaxies with similar \sratio\ colors, and it could alternatively indicate that the most extreme AGNs in the sample (in terms of line emission) are also associated with the most extreme star forming systems, or alternatively, that these \hii\ galaxies host a hidden AGN.

The best quadratic fit parameters to the relations involving \oi/FIR, \oii/FIR, \firmol, and \sfir\ are listed in Table~\ref{tab:scaling}.

\subsection{Nitrogen lines to FIR ratio}

The two nitrogen fine-structure transitions included in this study are the \niii\ and \nii\ lines. Both of these lines arise from the ionized gas (the ionization potential of atomic nitrogen is 14.5~eV), and the \nii\ transition in particular is a powerful tracer of the low-ionization, diffuse ionized gas \citep[e.g.,][]{rhc_goldsmith15,rhc_rhc16}.

\begin{figure*}
\begin{center}
\includegraphics[scale=0.23]{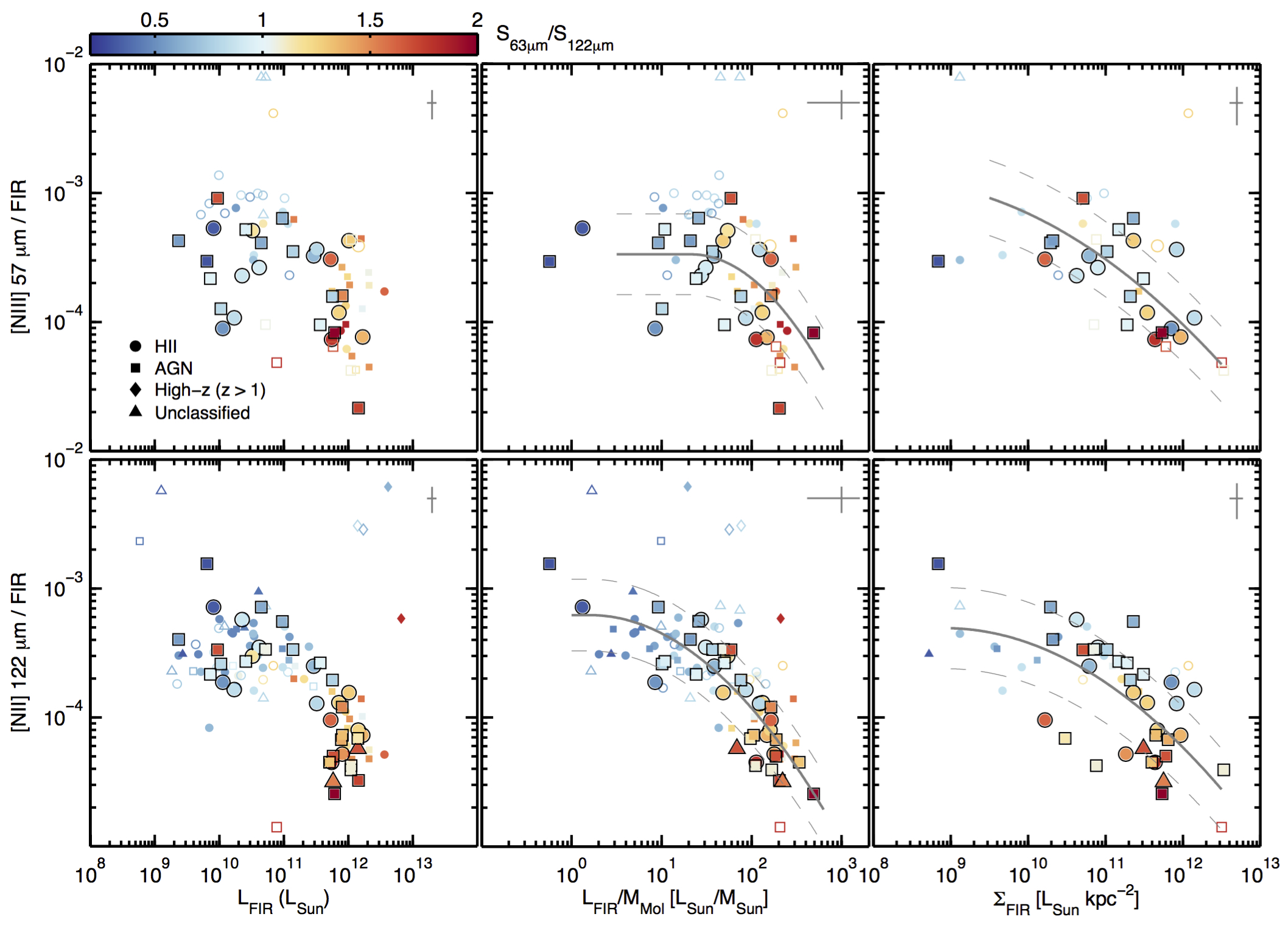}
\caption{Global \niii\ (top) and \nii\ (bottom) line to FIR continuum ratio for different types of galaxies  and high-$z$ ($z>1$) systems as a function of far-infrared luminosity (left), the ratio between far infrared luminosity and molecular gas mass (center), and infrared luminosity surface brightness (right). SHINING observations are marked with a black border. Open symbols indicate 3-$\sigma$ upper limits to the line flux and the color scale indicates the \sratio\ IR color. The gray lines show a simple quadratic fit to the data. (The best fit parameters are listed in Table~\ref{tab:scaling}. We set the fits to the maximum value of the quadratic equation below the value of \firmol\ or \sfir\ at which the maximum is reached.) Typical errorbars are plotted in the upper-right corner.}\label{Fig:n-deficits}
\end{center}
\end{figure*}

Figure~\ref{Fig:n-deficits} shows the \niii\ and \nii\ to FIR ratios as a function of \lfir, \firmol, and \sfir. The color scale indicates the \sratio\ color, and different markers are used to differentiate between \hii, AGN, and high-$z$ galaxies. We do not find a strong trend of decreasing \niii/FIR ratio as a function of \lfir\ (the Kendall $\tau$ coefficient is $\tau=-0.36$ with $p<0.01$). Although the average \niii/FIR decreases with \lfir, galaxies that have line ratios $\lesssim10^{-4}$ span the entire range of FIR luminosities sampled. We observe only weak trends between the \niii/FIR ratio and \firmol\ and \sfir\ (the Kendall $\tau$ coefficients for \firmol\ and \sfir\ are $\tau=-0.39$ with $p<0.01$, and $\tau=-0.49$ with $p<0.01$, respectively). In these cases, however, all the galaxies that have line ratios $\lesssim10^{-4}$ are confined to the bright end of the \sfir\ distribution.

In the case of the \nii\ transition, we find a clear trend of decreasing \nii/FIR ratio for all galaxy types as a function of \lfir\ ($\tau=-0.49$ with $p<0.01$), \firmol\ ($\tau=-0.64$ with $p<0.01$), \sfir\ ($\tau=-0.52$ with $p<0.01$), and \sratio\ color ($\tau=-0.56$ with $p<0.01$). This behavior could be driven by at least two effects. The first one is the low critical density of the \nii\ line, which at $T\approx8,000$~K is about $n_{\rm crit}\approx300$~cm$^{-3}$ \citep{rhc_hudson04}. In resolved observations of nearby galaxies, \cite{rhc_rhc16} find a trend of increasing electron densities with \sfir, reaching densities close to the critical density of the \nii\ line around $S_{70~\mu m}/S_{100~\mu m}\sim1.5$. These threshold values are consistent with those where we start observing the decline of the \nii/FIR ratio in our sample of \hii\ and AGN galaxies. The second factor that can drive the \nii/FIR ratio down is the increase in the N$^{++}$/N$^{+}$ ratio as the hardness of the ionizing radiation field increases. This effect could be important in (U)LIRGs, which show signatures of intense radiation field such as weakened Polycyclic Aromatic Hydrocarbon (PAH) emission \citep{rhc_genzel98,rhc_lutz98,rhc_laurent00,rhc_veilleux09,rhc_stierwalt13}, and high \oiii/\nii\ ratios \citep[][recall that the energy required to create O$^{++}$ ions is only $\sim$5~eV higher than that require to create N$^{++}$.]{rhc_zhao13}.

As with the other lines, the resulting best quadratic fit parameters to the relations involving \nii/FIR, \firmol, and \sfir\ are listed in Table~\ref{tab:scaling}.

\begin{figure*}
\begin{center}
\includegraphics[scale=0.23]{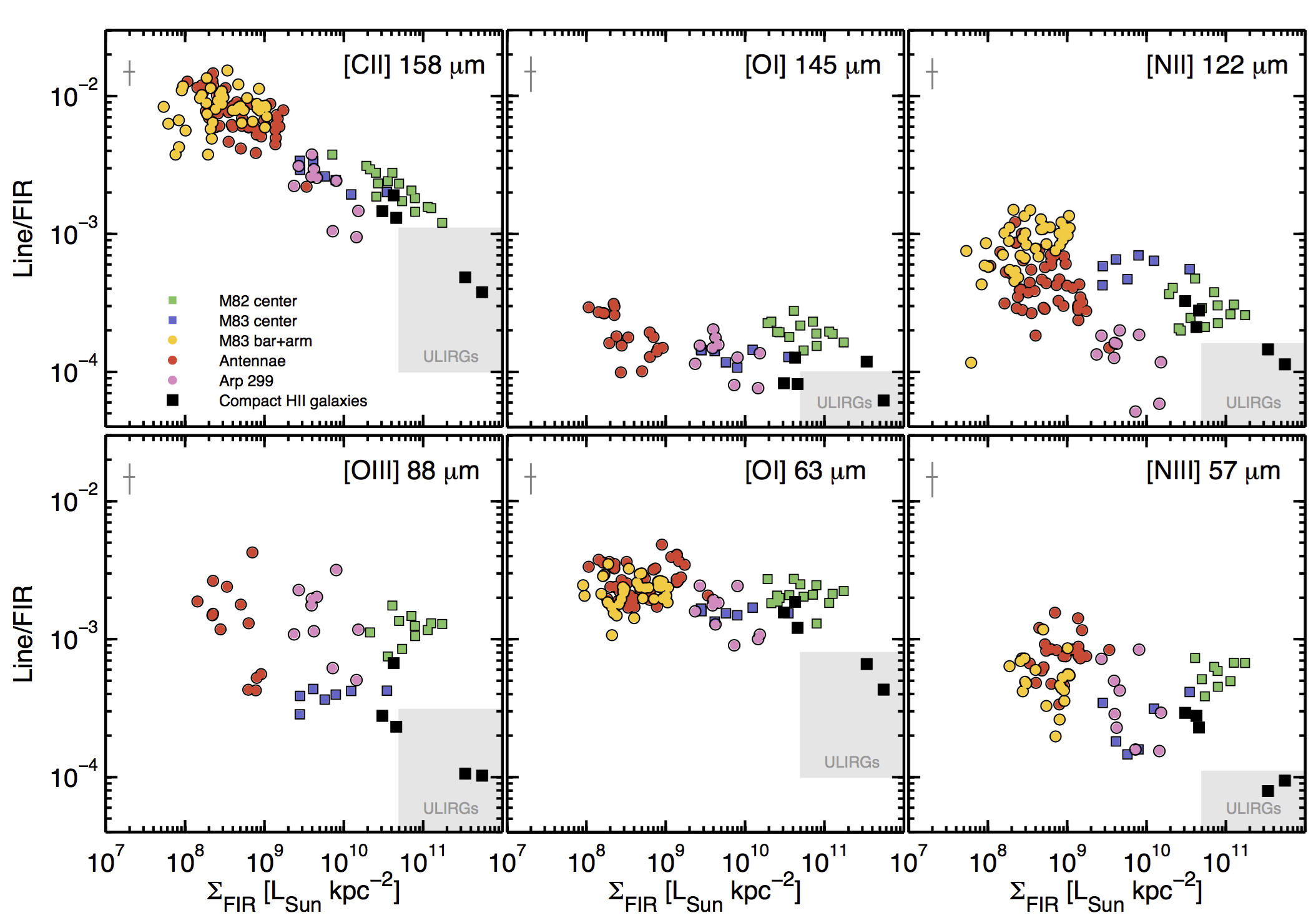}
\caption{Line to FIR continuum ratios in the SHINING \hii\ galaxies which are extended for the PACS spectrometer. Each point represents a different spaxel (size $9.4\arcsec$) and the spaxels centered on the star-forming nuclei are indicated with a square. The central spaxels from compact \hii\ galaxies are shown as black squares. The gray areas highlight the ranges of line to continuum ratios and \sfir\ typically observed in local ULIRGs with strong line deficits. Typical errorbars are plotted in the upper-left corner.}\label{Fig:maps-HII}
\end{center}
\end{figure*}

\subsection{Spatially resolved information \label{sec:resolved information}}

As we mentioned before, around 25 SHINING sources are nearby and thus extended for the PACS spectrometer spatial resolution (see Section~\ref{subsec:extended sources}). This gives us the opportunity to study the distribution of the line and continuum emission in active and star forming galaxies on kiloparsec and sub-kiloparsec scales. 

\subsubsection{\hii\ galaxies}

Figure~\ref{Fig:maps-HII} shows the spatially resolved line to FIR continuum ratios for the SHINING sample of extended \hii\ galaxies as a function of FIR surface brightness. Each color point in the figure represents an individual PACS spatial pixel. We also plot the central pixels from the sample of compact \hii\ galaxies (black squares). Line to FIR data are only displayed for spaxels in which the continuum at 63\,$\mu$m and 122\,$\mu$m are detected with a signal to noise ratio $\geq$5. 

The far-infrared continuum fluxes are calculated following:

\begin{equation}\label{eq:fira}
\begin{split}
\mathrm{FIR}(42.5\,\mu\mathrm{m}-122.5\,\mu\mathrm{m}) = 7.98 \times 10^{-15} \\
\times (4.67\ S_{63\,\mu\mathrm{m}} + S_{122\,\mu\mathrm{m}}),
\end{split}
\end{equation}

\noindent where FIR is in W\,m$^{-2}$, and \sblue\ and \sred\ are in Janskys. This definition is equivalent to the FIR definition adopted in Equation~\ref{eq:fir}, but uses the \sblue\ and \sred\ flux densities, which can be directly measured in the PACS maps from the \oi\ and \nii\ observations. Equation~\ref{eq:fira} was derived based on extrapolated \sblue\ and \sred\ continuum flux densities obtained  from fitting a single temperature greybody to the 60 and 100~$\mu$m continuum flux densities. The comparison between the resulting FIR fluxes using Equations~(\ref{eq:fir}) and (\ref{eq:fira}) is shown in Appendix~\ref{App:FIR}, Figure~\ref{Fig:FIRcomp}. The 1-$\sigma$ dispersion of the residual between the two FIR continuum flux definitions is only 0.026~dex.

On average, the relative intensity of the \cii\ line decreases with increasing \sfir\ (the Kendall $\tau$ coefficient is $\tau=-0.55$ with $p<0.01$). This result is similar to the observed galaxy integrated trends, but in this case it applies to regions with spatial scales between $\sim0.2-2$\,kpc. All \hii\ galaxies for which we have spatial information follow this trend. The \nii$/$FIR ratio is also reduced with increasing \sfir, although the trend is only clear individually for Arp~299 and the Antennae system ($\tau=0.47$ with $p<0.01$ for these two systems combined). With the exception of two central spaxels from the sample of compact \hii\ galaxies, the \oi$/$FIR and \oii$/$FIR ratios are relatively constant, with very little scatter ($\tau=-0.20$ and $\tau=-0.18$, respectively). On the other hand, the \oiii$/$FIR and \niii$/$FIR ratios show a large range of variation, both between different galaxies and for pixels in the same galaxy. This variation does not depend on the \sfir\ and is probably reflecting the stronger dependence of the \oiii\ and \niii\ line emission on the hardness of the radiation field. Dust extinction, or differences in the location where the line and the FIR continuum arise \citep[e.g.,][]{rhc_diaz-santos17}, could also contribute to the observed high dispersion.

Few spaxels in the sample of extended \hii\ galaxies show the line to continuum ratios observed in local ULIRGs with strong line deficits. The FIR surface brightness of various pixels in the central region of M~82 are comparable to the global values observed in ULIRGs, but their line emission is stronger. This indicates that even the most extreme star-forming regions in these galaxies are not able to reproduce the conditions of the interstellar medium in ULIRGs.

The central pixels from the sample of compact \hii\ galaxies have higher \sfir\ values and generally lower line to continuum ratios than the sample of extended galaxies. Of particular interest are the two pixels with the lowest line to continuum ratios, that coincide spatially with the position of the nuclear starbursts in NGC~253 and NGC~4945 and have far-infrared spectral properties similar to ULIRGs. Like many ULIRGs too, they have $L_{\mathrm{FIR}}/M_{\mathrm{mol}} \sim 250\,L_{\odot}\,M_{\odot}^{-1}$ and high molecular gas and SFR surface densities \citep[$\Sigma_{\mathrm{mol}} \sim 10^{3.5}\,M_{\odot}$\,pc$^{-2}$, $\Sigma_{\mathrm{SFR}} \sim 10^{1.5}\,M_{\odot}$\,kpc$^{-2}$;][]{rhc_chou07,rhc_weiss08,rhc_sakamoto11}. All this seems to suggest that we can interpret the ULIRG phenomenon as a scaled up version (by two orders of magnitude in terms of molecular gas mass and SFR) of the nuclear starbursts in NGC~4945 and NGC~253 \citep[note however that the nuclear dust temperature in these two galaxies is lower than in ULIRGs; ][]{rhc_gonzalez-alfonso15}.

\begin{figure}
\begin{center}
\includegraphics[scale=0.24]{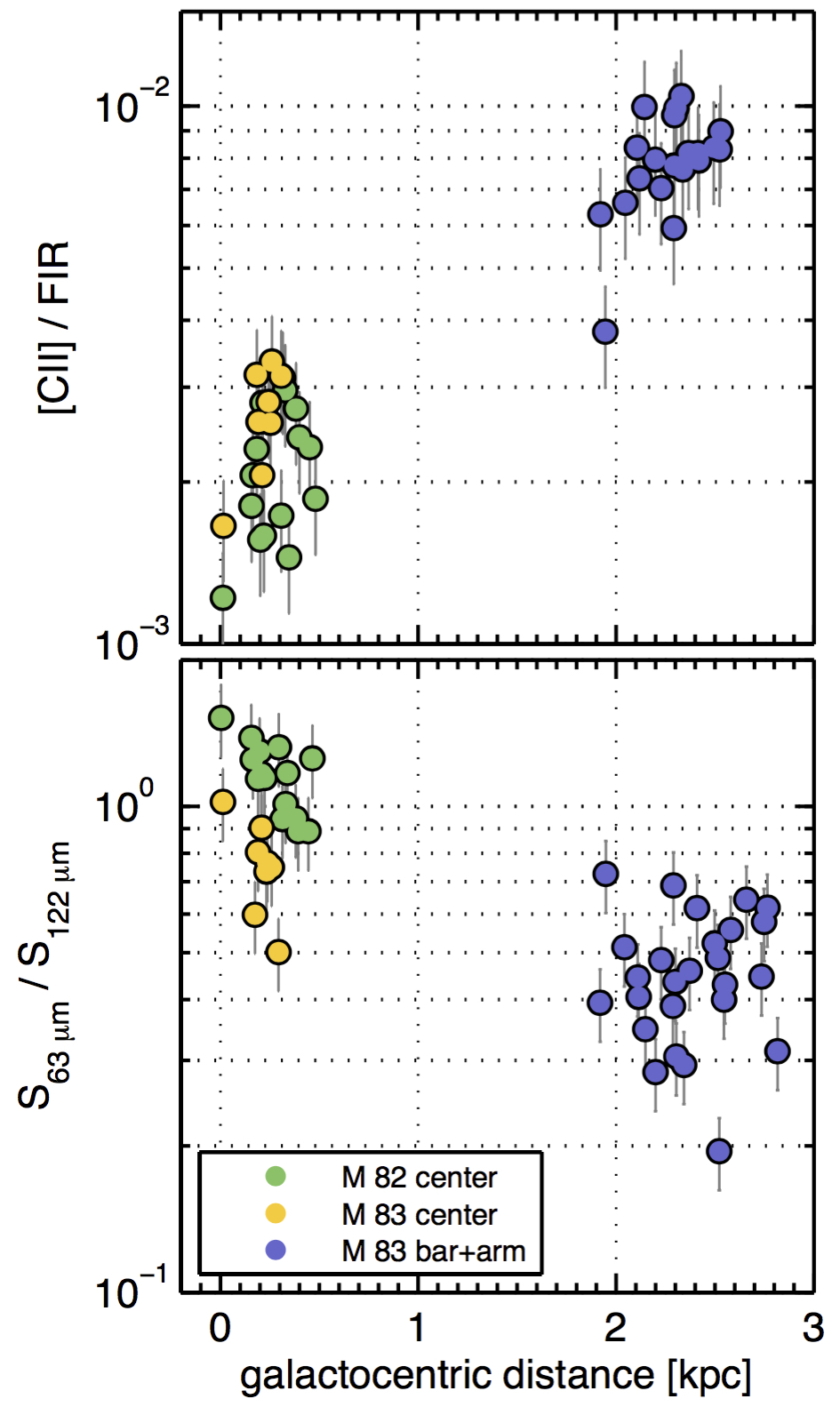}
\caption{\cii/FIR and \sratio\ as a function of the pixel distance to the galaxy center. The Antennae and Arp 299 are not shown in this figure because they are both interacting systems and, therefore, do not have a clearly defined galaxy center.}\label{Fig:galdist}
\end{center}
\end{figure}

The spatial scales probed in the maps of the closest galaxies M~82 and M~83 allows us to study the spatial distribution of their line to continuum ratios. Figure~\ref{Fig:galdist} shows that the lowest \cii$/$FIR ratios and highest \sratio\ colors in these galaxies are found at their centers. Similar trends are observed in M~33 \citep{rhc_kramer13}, M~31 \citep{rhc_kapala15}, and other nearby galaxies in the KINGFISH sample \citep{rhc_smith17}.

\begin{figure*}
\begin{center}
\includegraphics[scale=0.23]{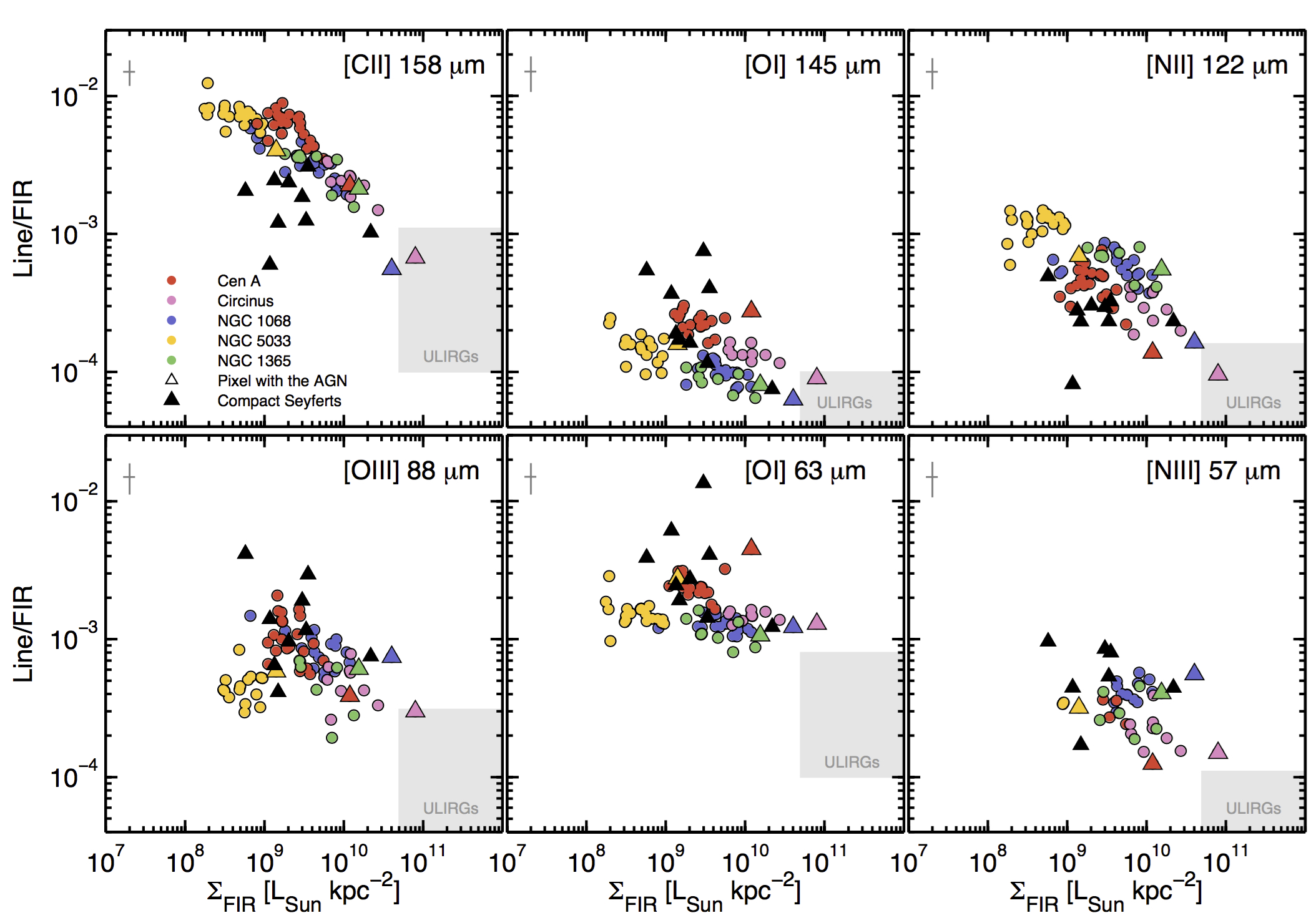}
\caption{Line to FIR continuum ratios in the SHINING AGN galaxies which are extended for the PACS spectrometer. Each point represents a different spaxel (size $9.4\arcsec$), and the spaxels centered on the AGN are indicated with a triangle. We also include the central spaxels from compact AGN galaxies shown as black triangles. The gray areas highlight the ranges of line to continuum ratios and \sfir\ typically observed in local ULIRGs with strong line deficits. Typical errorbars are plotted in the upper-left corner.}\label{Fig:maps-AGN}
\end{center}
\end{figure*}

\begin{figure*}
\begin{center}
\includegraphics[scale=0.35]{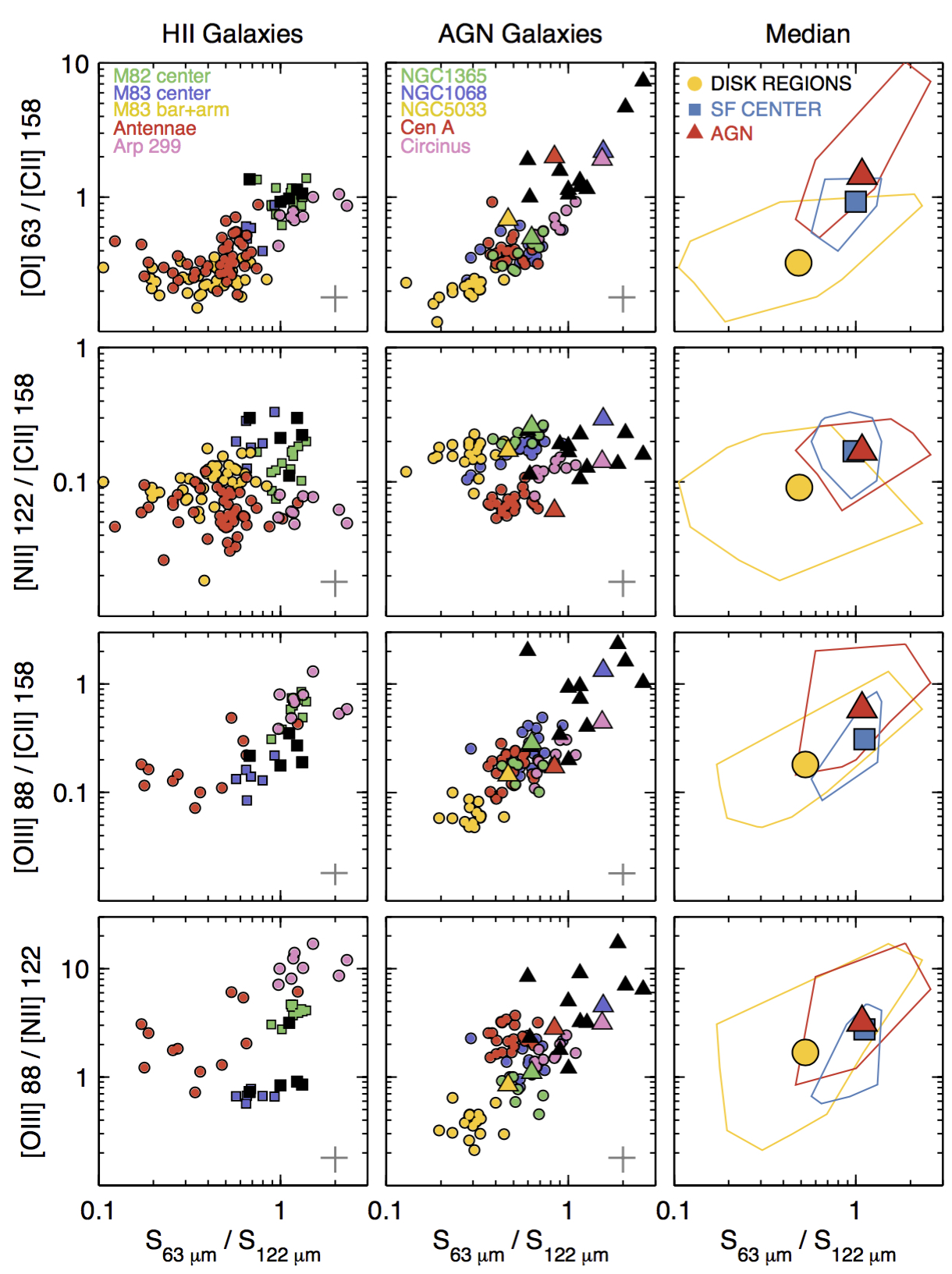}
\caption{Different line ratios as a function of \sratio\ IR color in the SHINING \hii\ (left) and AGN (center) galaxies which are extended for the PACS spectrometer. Each point represents a different spaxel (size $\sim9\arcsec$, spatial scales ranging between $\sim0.2-2$~kpc), and the spaxels centered on the star-forming center or the AGN are indicated with squares and triangles, respectively. The central spaxels from compact \hii\ and AGN galaxies are shown as black squares and triangles, respectively. The right panel shows the median and the range of values covered by regions in the disk (yellow lines and circles), central star-forming spaxels (blue lines and squares), and central AGN spaxels (red lines and triangles). Typical errorbars are plotted in the lower-right corner.}\label{Fig:maps-lr}
\end{center}
\end{figure*}

\begin{figure}
\begin{center}
\includegraphics[scale=0.13]{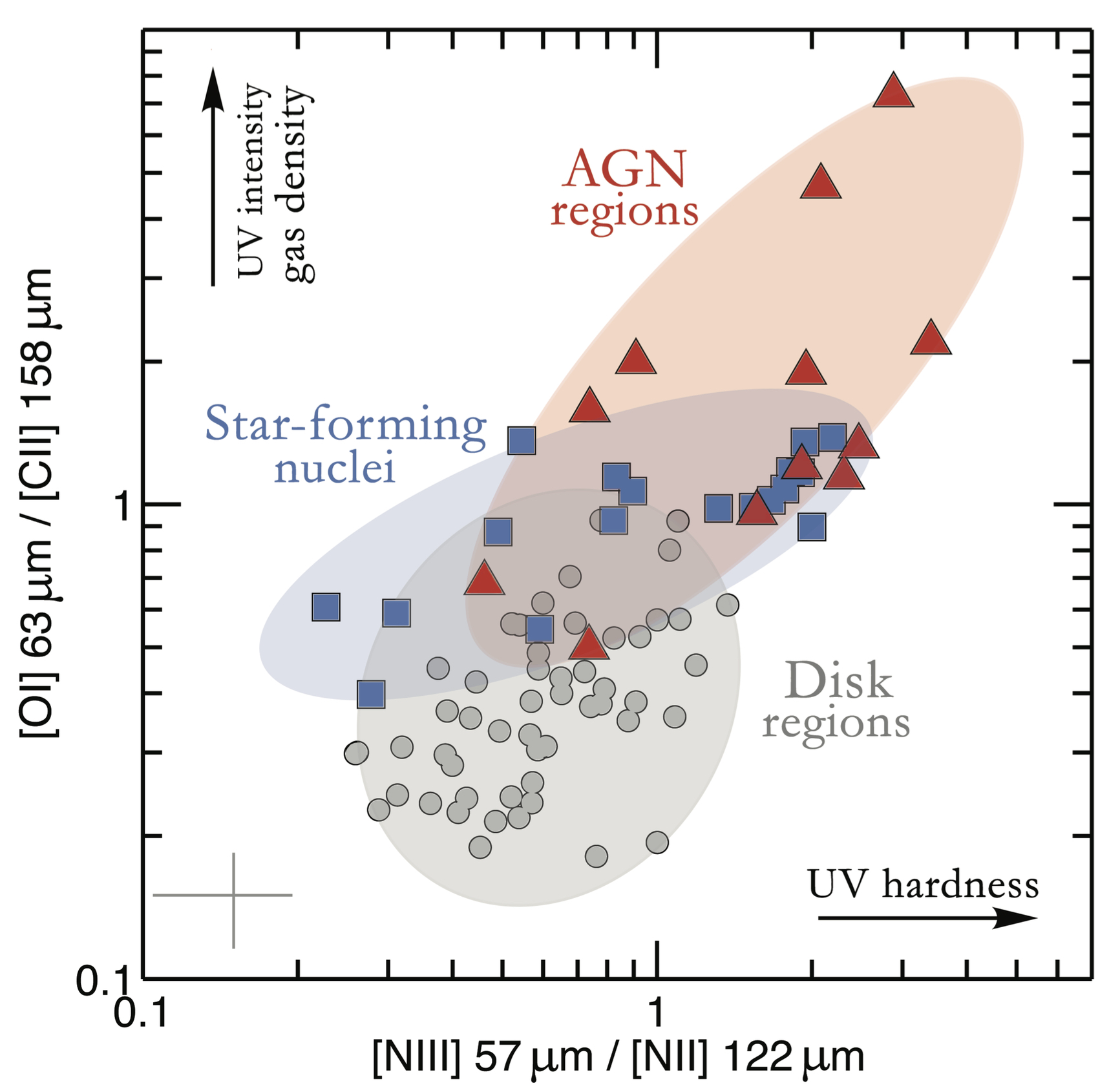}
\caption{\oi/\cii\ versus \niii/\nii\ line ratio for resolved regions in SHINING galaxies. We highlight AGN regions (red triangles), star-forming centers (blue squares), and disk regions (gray). The \oi/\cii\ line ratio increases with density of the gas and radiation field strength, while the \niii/\nii\ line ratio is sensitive to the hardness of the UV radiation field. The shaded boxes delineate regions distinguished by AGN, star-forming nuclei, and disk regions. Only central $\sim$kiloparsec size regions that host an AGN populate the top-right corner of the parameter space characterized by high density and/or radiation field strength and harsh radiation fields. A typical errorbar is plotted in the lower-left corner.}\label{fig:diagnostic}
\end{center}
\end{figure}

\subsubsection{AGN galaxies}

We can repeat the same kind of analysis with the line and continuum maps from the SHINING sample of extended Seyfert galaxies. In Figure~\ref{Fig:maps-AGN} we use different symbols to identify the spaxels containing the AGN (triangles) from those covering the galaxy disk (circles). This allows us to study the effects of AGN radiation and mechanical feedback on the line and continuum emission. We should note however that the physical sizes covered by the PACS spectrometer pixels are $\sim0.2-0.8$\,kpc for the luminosity distances of the studied galaxies, and therefore some contamination from nuclear star formation may be present.  

As we did in Figure~\ref{Fig:maps-HII} for the compact \hii\ galaxies, we include in Figure~\ref{Fig:maps-AGN} the AGN pixels from the sample of Seyfert galaxies with compact emission (black triangles). Pixel sizes for these galaxies range from 0.7 to 3.5\,kpc, therefore the detected emission may have a significant contribution coming from star-forming regions. 

We observe that, with the exception of the central regions in NGC1068, Circinus and NGC~4945, there is not a single spaxel in the studied sample of Seyfert galaxies with a line to continuum ratio as low as those observed in local ULIRGs and high-$z$ galaxies with strong line deficits. Extra-nuclear regions in Seyfert galaxies have line to continuum ratios, on average, only $\sim30\%$ higher than those observed in \hii\ galaxies, which is expected as the selection criteria for the Seyfert systems was based on their nuclear activity and did not take into account the properties of the host. This naturally led to select galaxies with normal star formation. The central AGN spaxels, on the other hand, tend to show weak \cii~158~$\mu$m but  particularly strong \niii, \oiii, and \oi\ line emission compared to the star-forming regions. In fact, the high \oi/FIR ratios observed in these systems could be used to identify the presence of AGN in galaxies when other methods cannot be applied \citep[note however that shocks can also produce strong \oi\ line emission; e.g.,][]{rhc_lutz03}. We discuss more in detail the AGN effect on the \cii\ and \oi\ line emission in Section~\ref{ir_color} and in Section 3.3 of Paper~II \citep{rhc_rhc18b}.

\subsubsection{Line ratios}\label{ir_color}

Given the variety of excitation conditions of the FIR cooling lines accessible with {\it Herschel}/PACS, their ratios can provide powerful diagnostics of the ISM physical conditions. Figure~\ref{Fig:maps-lr} shows various line ratios as a function of IR color \sratio\ in resolved regions ($\sim0.2-2$~kpc in size) of SHINING \hii\ and AGN galaxies. For galaxies that are classified as compact, we include the emission from the central spaxel (black squares for star-forming centers and triangles for AGNs).

\medskip 

\noindent {\bf \oi/\cii\ ratio:} For regions in \hii\ galaxies with \sratio~$\lesssim0.6$, we observe \oi/\cii\ line ratios in the $\sim0.2-0.5$ range, similar to the range of values observed in resolved regions of \hii\ galaxies in the KINGFISH sample \citep{rhc_rhc15}. For IR colors \sratio~$\gtrsim0.5$, we find a clear trend of increasing \oi/\cii\ ratios with increasing (warmer) IR colour in regions from both \hii\ and AGN galaxies (Kendall $\tau\approx0.5$ with $p<0.01$). The highest ratios are observed in the central spaxel of compact AGN systems. These trends are consistent with those observed in low-metallicity galaxies \citep{rhc_cormier15}, and a variety of other galaxy types presented by \cite{rhc_malhotra01} and \cite{rhc_brauher08}. The increase in the \oi/\cii\ line ratio with IR color --a proxy for the FUV heating intensity-- is expected as the \oi\ line is a more efficient coolant than the \cii\ line in warmer, denser gas.

\medskip 

\noindent {\bf \nii/\cii\ ratio:} For regions in both \hii\ and AGN galaxies we observe a relatively flat distribution of \nii/\cii\ ratios across the observed range of IR colors, although the scatter is large. In regions of \hii\ galaxies we measure a mean line ratio of $0.1\pm0.06$, while in regions in AGN galaxies we measure a mean line ratio of $0.14\pm0.05$. These values are consistent with those observed in the Milky Way \citep{rhc_goldsmith15}, and in the samples of galaxies of \cite{rhc_malhotra01} and \cite{rhc_brauher08}. 

Given that the \nii\ line emission can only arise in ionized gas, the \nii\ to \cii\ line ratio can be used to constrain the fraction of the \cii\ emission originating in the neutral gas, $f_{\rm [CII]}^{\rm neutral}$. This fraction can be calculated as follows:

\begin{equation}
f_{\rm [CII]}^{\rm neutral}=1-R_{\rm ion}\times\bigg(\frac{{\rm [NII]}~122~\mu m}{\rm [CII]~158~\mu m}\bigg)_{\rm observed},
\end{equation}

\noindent where $R_{\rm ion}$ is the expected \cii/\nii\ ratio if all the \cii\ emission originates in the ionized gas phase. $R_{\rm ion}$ is a function of the electron density ($n_{\rm e}$) of the ionized gas\footnote{In principle $R_{\rm ion}$ also depends on the ionization state of the \hii\ gas, which in turn is a function of the spectrum of the stellar radiation and the ionization parameter. For instance, in dense \hii\ regions we expect nitrogen to be preferentially doubly ionized \citep[e.g.,][]{rhc_abel09}.}, so in order to estimate $f_{\rm [CII]}^{\rm neutral}$ first we need to constrain the range of electron densities in our sample. For this we use the scaling relation between electron density and IR color in \cite{rhc_rhc16}. We find that our regions roughly span the range of densities $n_{\rm e}\sim10-200$~cm$^{-3}$. In this range, and assuming the collision rates of \cite{rhc_tayal08,rhc_tayal11} and Galactic gas phase abundances for carbon and nitrogen \citep{rhc_meyer97,rhc_sofia04}, we find that $R_{\rm ion}$ varies from 
$R_{\rm ion}(n_{\rm}=10~{\rm cm}^{-3})=4$ to $R_{\rm ion}(n_{\rm}=200~{\rm cm}^{-3})=0.9$. For the roughly constant \nii/\cii\ ratio of $\approx0.1$, this implies that the contribution to the \cii\ emission from neutral gas in our sample of resolved regions increases as a function of \sratio\ from $f_{\rm [CII]}^{\rm neutral}\approx60\%$ to 90\%. 

\medskip 

\noindent {\bf \oiii/\cii\ ratio:} The \oiii\ line emission originates in the narrow line region of AGNs or \hii\ regions where O and B stars produce photons with energies higher than the 35~eV required to create O$^{++}$. In both groups of resolved regions we observe a trend of increasing \oiii/\cii\ line ratios as a function of increasing IR color (Kendall $\tau\approx0.5$ with $p\approx0.01$), but the scatter is large. This is in contrast to the roughly constant ratio we find when we compare the \cii\ and \nii\ line emission. This could be an indication that the contribution to the \cii\ emission from the ionized phase is dominated by low-excitation N$^{+}$ and O$^{+}$ gas, rather than highly ionized O$^{++}$ regions \citep{rhc_brauher08}.

\medskip 

\noindent {\bf \oiii/\nii\ ratio:} For resolved regions in \hii\ galaxies we observe a tentative trend of increasing \oiii/\nii\ line ratios as a function of IR color (Kendall $\tau=0.37$ with $p<0.01$). What is interesting is that regions in M~83 and Arp~299 that have similar IR colors can have \oiii/\nii\ line ratios that differ by an order of magnitude. In \hii\ regions, the \oiii/\nii\ line ratio is relatively insensitive to ionized gas density (the critical densities of the \oiii\ and \nii\ lines are 510 and 310~cm$^{-3}$, respectively), but very sensitive to the effective temperature $T_{\rm eff}$ of the ionizing source. This line ratio also depends on the ionization parameter and the N/O abundance. Based on the models of \cite{rhc_rubin85} \citep[see also][]{rhc_ferkinhoff11}, we find stellar effective temperatures that increase from approximately $3.4\times10^4$~K in M~83 to $3.7\times10^4$~K in Arp~299, which in turn corresponds to stellar classifications of O9.5 to O8.5 for the most luminous stars  \citep[following the classification of][]{rhc_vacca96,rhc_sternberg03}. 

In the central regions of AGN galaxies we observe a trend of increasing \oiii/\nii\ line ratios with \sratio\ (Kendall $\tau=0.42$ with $p<0.01$). In the narrow line region (NLR) of AGNs, the \oiii/\nii\ line ratio is a sensitive indicator of the ionization parameter $U$. Using the radiation pressure-dominated photoionization model of \cite{rhc_groves04}, we find ionization parameters that vary from $U=10^{-3.5}$ in NGC~5033 to $U=10^{-3}$ in NGC~1068.

\subsubsection{\oi/\cii\ versus \niii/\nii: a potential line diagnostic of AGN and star-forming regions}\label{diagnostic}

Another useful far-infrared line diagnostic that can help to discriminate between AGN and star formation activity is \oi/\cii\ versus \niii/\nii. This plot tracks the density and radiation field strength on the y-axis, while the x-axis is sensitive to the hardness of the UV radiation field.  Figure~\ref{fig:diagnostic} shows the position of $\sim$kiloparsec size AGN, star-forming nuclei, and disk regions in this diagnostic diagram. We observe that disk regions stand apart from nuclei that are powered by star formation and accretion-powered disks, and that the latter are the only ones that populate the top-right corner of the diagram ($\oi/\cii\gtrsim1.6$), characterized by harsh and intense UV radiation fields and high-density gas. This is consistent with the model results by \cite{rhc_abel09} \citep[see also][]{rhc_maloney96}, which predict that in the ionization parameter range $10^{-4}\lesssim U \lesssim 0.01$ the \oi/\cii\ line ratio increases from $\sim1$ to $\sim5$ in AGNs, and only from $\sim0.3$ to $\sim0.6$ in starbursts. A two-sample, two-dimensional Kolmogorov-Smirnov test \citep{rhc_peacock83} shows that AGN, star-forming nuclei, and disk region are distinct populations (the null hypothesis that there is no difference between these three populations is rejected with p-value $p<0.01$).

While there are other diagnostics based on mid-infrared lines that can help to discriminate more clearly between pure starbursts and AGNs \citep[e.g.,][]{rhc_sturm02,rhc_dale06,rhc_armus07,rhc_fo16}, the diagnostic diagram presented here can be useful to differentiate strong AGNs from star-forming dominated nuclei and extra-nuclear regions in the absence of tracers in the mid-infrared.

\section{Summary and Conclusions}\label{conclusions}

In this paper we present the main results of the analysis of the infrared fine-structure line emission in 52 galaxies that constitute the SHINING sample (PI Sturm). These galaxies range from star-forming to AGN dominated systems, including luminous and ultra-luminous infrared galaxies. Observations of the infrared lines were obtained using the PACS instrument on board {\it Herschel} and in this paper we focus on the six brightest \hii\ and PDR lines in the $55-210~\mu$m wavelength range. The lines are: \cii~158$\mu$m, \oii, \nii, \oiii, \oi, and \niii.

We highlight the following points:

\begin{enumerate}
\item{{\it Global line-to-continuum ratios.} We find that the fraction of galaxies in the local universe with strong line deficits increases with FIR luminosity. This is in good agreement with previous studies based on ISO and {\it Herschel} observations \citep{rhc_malhotra01,rhc_luhman03,rhc_brauher08,rhc_gracia-carpio11,rhc_farrah13,rhc_diaz-santos17}. High-$z$ starburst galaxies follow a similar trend, but shifted to higher $L_{\rm FIR}$ values \citep[e.g.,][]{rhc_maiolino09,rhc_stacey10,rhc_brisbin15}. If we analyze the line to continuum ratios as a function of \sfir\ or \firmol\ --instead of $L_{\rm FIR}$-- we find tighter correlations that help to reconcile the observed shift in the distribution of low and high-$z$ galaxies. In the particular case of the \cii\ line, the dispersion around the best quadratic fit to the \cii/FIR--\sfir\ relation is only 0.2~dex over almost five orders of magnitude in \sfir\ \citep[see also ][]{rhc_lutz16}. The reason is that \sfir\ or \firmol\ are galaxy properties more closely related to the physical parameters controlling the relative intensity of the line (e.g., \gof, gas density, $U$). Independent of redshift, galaxies with more extreme ISM properties such as $L_{\mathrm{FIR}}/M_{\mathrm{mol}} \gtrsim 80\,L_{\odot}\,M_{\odot}^{-1}$ and/or \sfir$\gtrsim10^{11}$~$L_{\odot}$~kpc$^{-2}$ exhibit low line to FIR continuum ratios \citep[see also][]{rhc_gracia-carpio11,rhc_lutz16}. In this context, it is wrong to assume that all LIRGs exhibit line deficits: for a fixed FIR luminosity we observe a clear trend of decreasing line to continuum ratios as LIRGs become more compact. In fact, LIRGs in our sample with infrared sizes $R_{70}\gtrsim1$~kpc have line to continuum ratios similar to those observed in normal, star-forming galaxies \citep[see also][]{rhc_diaz-santos14}.}

\item{{\it Spatially resolved line-to-continuum ratios.} For nearby star-forming and Seyfert galaxies that we can spatially resolve with PACS we find trends of line to continuum ratios as a function of \sratio\ (a proxy for the dust temperature) similar to those observed based on global values. Even though we have about 30 regions with \sratio\ infrared colors similar to those observed in ULIRGs, only in a few cases --the nuclear starbursts in Arp~299 and NGC~253, the AGN regions in NGC~1068 and Circinus-- we find ULIRG-like extremely low line-to-continuum ratios.}

\item{{\it Line ratios.} The wide range of excitation conditions probed by the infrared fine-structure lines can be used to infer physical properties of the gas. From the analysis of the line emission in resolved regions of star-forming and Seyfert galaxies we find that: 

\noindent (1) The fraction of \cii\ emission arising from the neutral gas increases from $\sim60\%$ in moderately star-forming environments to $\sim90\%$ in active star-forming regions. 

\noindent (2) The \oi/\cii\ ratios increase as a function of IR color, with the highest ratios observed in the central spaxel of the compact AGN galaxies. This is most likely a consequence of the higher densities and higher radiation fields expected in AGN and starburst regions, as the \oi\ line is a more efficient coolant of the warm, dense gas than \cii. 

\noindent (3)  The \cii\ emission emitted in the ionized phase is more associated with low-excitation, diffuse gas rather than highly-ionized dense \hii\ gas (as suggested by the flat distribution of the \nii/\cii\ ratio and the trend of increasing \oiii/\cii\ ratio as a function of infrared color). 

\noindent (4) From the analysis of the \oiii/\nii\ ratio, the effective temperature of the most luminous stars ionizing the gas in our star-forming regions vary from $T_{\rm eff} = 34,000$~K (O9.5) to $T_{\rm eff} = 37,000$~K (O8.5).

\noindent (5) In the diagnostic diagram based on the \oi/\cii\ and \niii/\nii\ ratios, we observe a transition from disk regions to star-forming nuclei to AGN regions as the hardness and intensity of the UV radiation field and the density of the neutral gas increases.
}

\end{enumerate}

\begin{acknowledgements}
We thank the referee for helpful and constructive comments that improved the paper. We thank Mark Wolfire, Natascha F{\"o}rster Schreiber, Shmuel Bialy, and Taro Shimizu for helpful discussions and comments. RHC would like to thank the support and encouragement from Fares Bravo Garrido and dedicates this paper with love to Fares and Olivia. Basic research in IR astronomy at NRL is funded by the US ONR.  JF also acknowledges support from the NHSC. E.GA is a Research Associate at the Harvard-Smithsonian Center for Astrophysics, and thanks the Spanish Ministerio de Econom\'{\i}a y Competitividad for support under projects FIS2012-39162-C06-01 and  ESP2015-65597-C4-1-R, and NASA grant ADAP NNX15AE56G. RM acknowledges the ERC Advanced Grant 695671 ``QUENCH'' and support from the Science and Technology Facilities Council (STFC). The Herschel spacecraft was designed, built, tested, and launched under a contract to ESA managed by the Herschel/Planck Project team by an industrial consortium under the overall responsibility of the prime contractor Thales Alenia Space (Cannes), and including Astrium (Friedrichshafen) responsible for the payload module and for system testing at spacecraft level, Thales Alenia Space (Turin) responsible for the service module, and Astrium (Toulouse) responsible for the telescope, with in excess of a hundred subcontractors. PACS has been developed by a consortium of institutes led by MPE (Germany) and including UVIE (Austria); KU Leuven, CSL, IMEC (Belgium); CEA, LAM (France); MPIA (Germany); INAF-IFSI/OAA/OAP/OAT, LENS, SISSA (Italy); IAC (Spain). This development has been supported by the funding agencies BMVIT (Austria), ESA-PRODEX (Belgium), CEA/CNES (France), DLR (Germany), ASI/INAF (Italy), and CICYT/MCYT (Spain). HCSS / HSpot / HIPE is a joint development (are joint developments) by the Herschel Science Ground Segment Consortium, consisting of ESA, the NASA Herschel Science Center, and the HIFI, PACS and SPIRE consortia. This publication makes use of data products from the Sloan Digital Sky Survey (SDSS).  Funding for the Sloan Digital Sky Survey has been provided by the Alfred P. Sloan Foundation, the Participating Institutions, the National Aeronautics and Space Administration, the National Science Foundation, the U.S. Department of Energy, the Japanese Monbukagakusho, and the Max Planck Society. This research has also made use of the NASA/IPAC Extragalactic Database (NED) which is operated by the Jet Propulsion Laboratory, California Institute of Technology, under contract with the National Aeronautics and Space Administration.
\end{acknowledgements}

\software{\textsc{CLOUDY} \citep{rhc_ferland13}, HIPE \citep[v13.0;][]{rhc_ott10}} 
\facility{Herschel \citep{rhc_pilbratt10}}


\appendix

\section{Comparison between FIR continuum fluxes} \label{App:FIR}

\begin{figure}
\begin{center}
\includegraphics[scale=0.15]{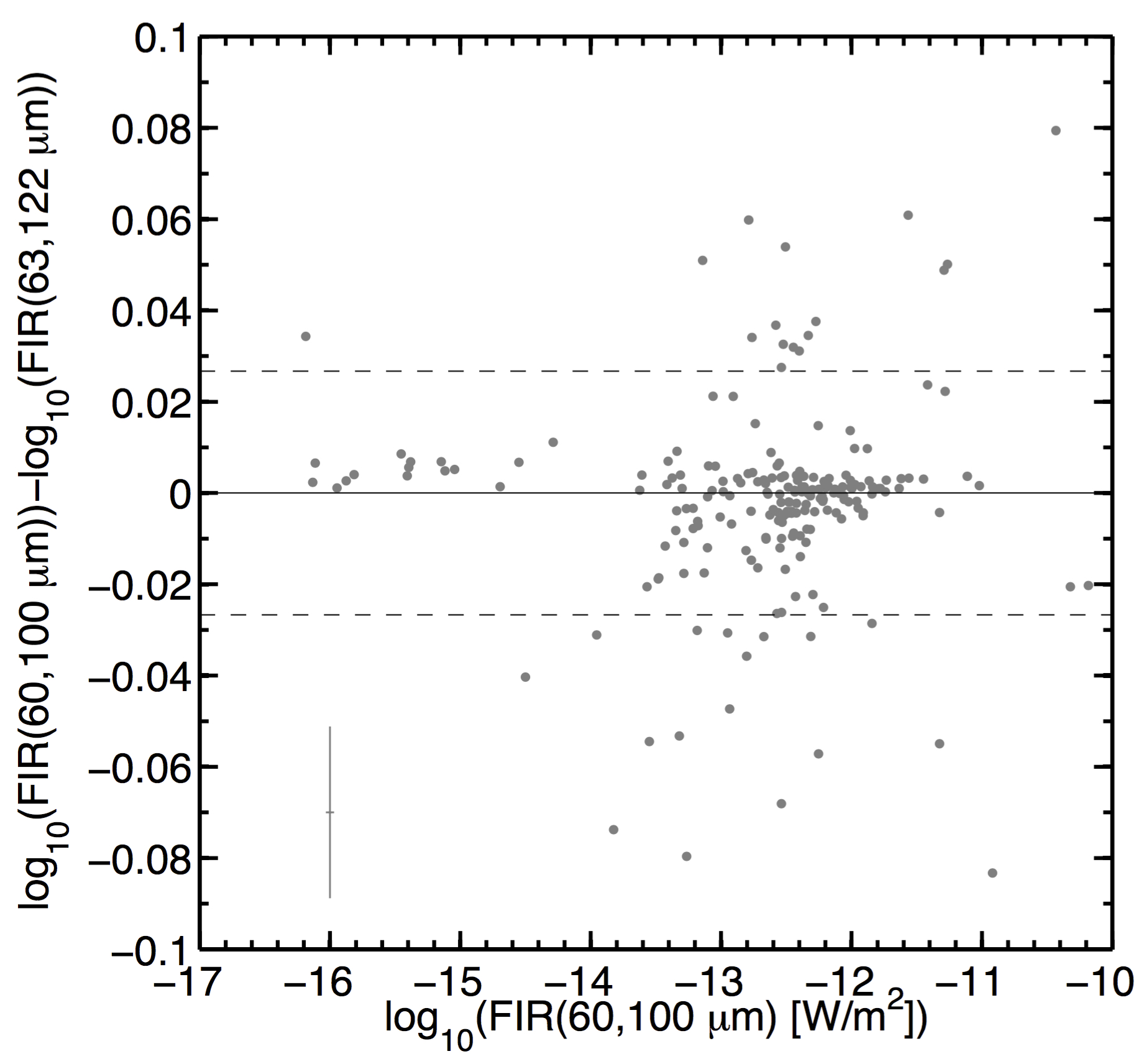}
\caption{Comparison between the FIR continuum fluxes calculated based on the continuum flux densities at 60 and 100~$\mu$m using Equation~(\ref{eq:fir}), and 63 and 122~$\mu$m using Equation~(\ref{eq:fira}). The 1-$\sigma$ standard deviation of the residual is 0.026~dex (dashed lines). A typical errorbar is plotted in the lower-left corner.
\label{Fig:FIRcomp}}
\end{center}
\end{figure}

\section{Summary of Scaling Parameters}

\floattable
\begin{table*}
\begin{center}
\caption{Summary of Scaling Parameters\label{tab:scaling}}
\begin{tabular}{c|cccccc}
\hline \hline
\multicolumn{7}{c}{${\rm log_{10}(Line/FIR)}=\alpha\times{\rm [log_{10}}(L_{\rm FIR}/M_{\rm mol})]^2+\beta\times{\rm log_{10}}(L_{\rm FIR}/M_{\rm mol})+\gamma$}\\
\hline
Line & $\alpha$ & $\beta$ & $\gamma$ & $L_{\rm FIR}/M_{\rm mol}$ at & Max. Ratio & 1$\sigma$ \\ 
 &  &  &  & Max. Ratio & $\times10^{-3}$ & dex \\ \hline
\hline
\niii\ & $-$0.7068 & 1.952 & $-$4.645 & 21.38 & 0.54 & 0.41 \\
\oi\ & $-$0.2346 & 0.387 & $-$2.826 & 6.60 & 2.15 & 0.31 \\
\oiii\ & $-$0.1459 & 2.653 & $-$14.768 & 5.50 & 1.97 & 0.42  \\
\nii\ & $-$0.0988 & 1.766 & $-$11.203 & 1.58 & 0.49 & 0.31 \\
\oii\ & $-$0.1347 & 2.619 & $-$16.428 & 3.46 & 0.20 & 0.29 \\
\cii~158~$\mu$m &  $-$0.1044 & 1.843 & $-$10.458 & 2.00 & 4.70 & 0.24 \\
\hline
\multicolumn{7}{c}{${\rm log_{10}(Line/FIR)}=\alpha^\prime\times{\rm [log_{10}}(\Sigma_{\rm FIR})]^2+\beta^\prime\times{\rm log_{10}}(\Sigma_{\rm FIR})+\gamma^\prime$}\\
\hline
Line & $\alpha^\prime$ & $\beta^\prime$ & $\gamma^\prime$ & log$_{10}(\Sigma_{\rm FIR})$ at & Max. Ratio & 1$\sigma$ \\ 
 &  &  &  & Max. Ratio & $\times10^{-3}$ & dex \\ \hline
\hline
\niii\ & $-$0.0771 & 1.296 & $-$8.447 & 9.50 & 0.81 & 0.30 \\
\oi\ &  $-$0.1225 & 2.280 & $-$13.126 & 9.30 & 3.05 & 0.30 \\
\oiii\ & $-$0.1723 & 0.254 & $-$3.047 & 9.10 & 1.11 & 0.51  \\
\nii\ & $-$0.2240 & 0.092 & $-$3.216 & 9.00 & 0.62 & 0.28 \\
\oii\ & $-$0.1454 & 0.156 & $-$3.819 & 9.71 & 0.17 & 0.26 \\
\cii~158~$\mu$m &  $-$0.2044 & 0.156 & $-$2.414 & 8.80 & 4.12 & 0.29 \\
\hline
\end{tabular}
\end{center} \label{quad_fit}
\end{table*}

\section{Global flux measurements for the SHINING galaxies}

\floattable
\begin{deluxetable}{lcccccccc}
\rotate
\tablecaption{Integrated line fluxes for galaxies in the SHINING sample \label{table:Global measurements}}
\tablehead{
\colhead{Source} & \colhead{\niii} & \colhead{\oi} & \colhead{\oiii} & \colhead{\nii} & \colhead{\oii} & \colhead{\cii~158~$\mu$m} & \colhead{$S_{63}$} & \colhead{$S_{122}$} \\
& \colhead{W~m$^{-2}$} & \colhead{W~m$^{-2}$} & \colhead{W~m$^{-2}$} & \colhead{W~m$^{-2}$} & \colhead{W~m$^{-2}$} & \colhead{W~m$^{-2}$} & \colhead{Jy} & \colhead{Jy}}
\startdata  
M82 Center	&	3.51E-14	$\pm$	1.10E-15	&	1.51E-13	$\pm$	5.20E-15	&	8.31E-14	$\pm$	2.60E-15	&	2.05E-14	$\pm$	1.20E-16	&	1.38E-14	$\pm$	1.20E-16	&	1.43E-13	$\pm$	1.00E-15	&	1570	$\pm$	94.20	&	1350	$\pm$	81.00	\\
M82 S outflow	&			\ldots	&			\ldots	&			\ldots	&	1.67E-15	$\pm$	1.20E-17	&			\ldots	&			\ldots	&			\ldots	&			\ldots	\\
M82 N outflow	&			\ldots	&			\ldots	&			\ldots	&	1.07E-15	$\pm$	2.10E-17	&			\ldots	&			\ldots	&			\ldots	&			\ldots	\\
M83 Center	&	1.90E-15	$\pm$	1.20E-16	&	9.85E-15	$\pm$	5.60E-17	&	2.45E-15	$\pm$	5.00E-17	&	3.35E-15	$\pm$	2.20E-17	&	7.30E-16	$\pm$	2.70E-17	&	1.72E-14	$\pm$	2.40E-16	&	127.33	$\pm$	7.64	&	156.24	$\pm$	9.37	\\
M83 Bar-spiral arm	&	6.15E-16	$\pm$	5.00E-17	&	2.73E-15	$\pm$	6.10E-17	&			\ldots	&	1.29E-15	$\pm$	1.50E-17	&			\ldots	&	9.26E-15	$\pm$	1.20E-16	&	21.92	$\pm$	1.32	&	50.02	$\pm$	3.00	\\
M83 Eastern arm	&	1.15E-16	$\pm$	1.50E-17	&	7.14E-16	$\pm$	2.20E-17	&			\ldots	&	2.84E-16	$\pm$	7.50E-18	&			\ldots	&	2.64E-15	$\pm$	7.70E-17	&	4.85	$\pm$	0.29	&	15.38	$\pm$	0.92	\\
NGC 253	&	5.43E-15	$\pm$	1.40E-16	&	3.38E-14	$\pm$	4.40E-16	&	7.55E-15	$\pm$	1.40E-16	&	8.28E-15	$\pm$	7.10E-17	&	3.96E-15	$\pm$	1.40E-16	&	4.56E-14	$\pm$	1.90E-16	&	1001	$\pm$	60.06	&	1212	$\pm$	72.72	\\
NGC 1808	&	1.15E-15	$\pm$	6.00E-17	&	9.61E-15	$\pm$	2.20E-16	&	2.39E-15	$\pm$	8.20E-17	&	2.88E-15	$\pm$	1.50E-17	&	6.38E-16	$\pm$	2.20E-17	&	1.67E-14	$\pm$	1.20E-16	&	127.05	$\pm$	7.62	&	144.2	$\pm$	8.65	\\
NGC 3256	&	1.57E-15	$\pm$	1.30E-16	&	9.43E-15	$\pm$	9.50E-17	&	4.37E-15	$\pm$	5.90E-17	&	1.21E-15	$\pm$	1.40E-17	&	8.44E-16	$\pm$	2.00E-17	&	1.79E-14	$\pm$	7.80E-17	&	99.84	$\pm$	5.99	&	148.83	$\pm$	8.93	\\
NGC 4038	&	2.87E-16	$\pm$	5.60E-17	&	1.61E-15	$\pm$	6.00E-17	&	8.01E-16	$\pm$	3.10E-17	&	3.86E-16	$\pm$	5.50E-18	&	1.07E-16	$\pm$	4.10E-18	&	4.15E-15	$\pm$	2.80E-17	&	10.01	$\pm$	0.60	&	29.89	$\pm$	1.79	\\
NGC 4039	&	8.04E-16	$\pm$	3.40E-17	&	3.36E-15	$\pm$	7.70E-17	&			\ldots	&	2.88E-16	$\pm$	7.70E-18	&			\ldots	&	6.20E-15	$\pm$	6.40E-17	&	18.29	$\pm$	1.10	&	40.59	$\pm$	2.44	\\
Overlap	&	6.88E-16	$\pm$	4.80E-17	&	2.38E-15	$\pm$	6.90E-17	&			\ldots	&	3.35E-16	$\pm$	1.60E-17	&			\ldots	&	6.97E-15	$\pm$	4.00E-17	&	13.12	$\pm$	0.79	&	23.98	$\pm$	1.44	\\
NGC 4945	&	2.82E-15	$\pm$	1.30E-16	&	2.67E-14	$\pm$	5.00E-15	&	3.78E-15	$\pm$	1.10E-16	&	5.91E-15	$\pm$	2.00E-16	&	3.94E-15	$\pm$	3.60E-17	&	3.17E-14	$\pm$	3.20E-16	&	573.8	$\pm$	34.43	&	1079.1	$\pm$	64.75	\\
NGC 7552	&	9.45E-16	$\pm$	1.70E-16	&	4.79E-15	$\pm$	4.30E-17	&	1.08E-15	$\pm$	4.00E-17	&	1.25E-15	$\pm$	1.30E-17	&	3.35E-16	$\pm$	1.20E-17	&	7.99E-15	$\pm$	4.50E-17	&	85.57	$\pm$	5.13	&	92.15	$\pm$	5.53	\\
NGC 1365	&	1.88E-15	$\pm$	1.50E-16	&	5.69E-15	$\pm$	1.60E-16	&	2.60E-15	$\pm$	5.80E-17	&	3.29E-15	$\pm$	2.20E-17	&	4.20E-16	$\pm$	1.30E-17	&	1.35E-14	$\pm$	1.20E-16	&	111.48	$\pm$	6.69	&	183.54	$\pm$	11.01	\\
NGC 3783	&	5.18E-17	$\pm$	1.70E-17	&	6.33E-16	$\pm$	4.00E-17	&	1.53E-16	$\pm$	1.30E-17	&	8.95E-17	$\pm$	8.70E-18	&	3.03E-17	$\pm$	4.50E-18	&	6.88E-16	$\pm$	9.90E-18	&	2.34	$\pm$	0.14	&	5.88	$\pm$	0.35	\\
NGC 4051	&	$<$2.55E-17			&	3.43E-16	$\pm$	2.50E-17	&	7.50E-17	$\pm$	1.50E-17	&	4.96E-17	$\pm$	5.30E-18	&	2.22E-17	$\pm$	3.60E-18	&	6.11E-16	$\pm$	9.20E-18	&	2.48	$\pm$	0.15	&	7.64	$\pm$	0.46	\\
NGC 4151	&	2.19E-16	$\pm$	3.40E-17	&	2.92E-15	$\pm$	6.50E-17	&	4.64E-16	$\pm$	3.20E-17	&	8.63E-17	$\pm$	9.60E-18	&	1.79E-16	$\pm$	5.80E-18	&	7.26E-16	$\pm$	1.30E-17	&	6.84	$\pm$	0.41	&	5.8	$\pm$	0.35	\\
NGC 4593	&	1.84E-17	$\pm$	5.90E-18	&	1.97E-16	$\pm$	2.20E-17	&	5.51E-17	$\pm$	8.10E-18	&	3.77E-17	$\pm$	8.60E-18	&	2.32E-17	$\pm$	4.00E-18	&	2.19E-16	$\pm$	1.10E-17	&	3.44	$\pm$	0.21	&	4.66	$\pm$	0.28	\\
NGC 5033	&	2.11E-16	$\pm$	4.00E-17	&	1.67E-15	$\pm$	4.40E-17	&	4.95E-16	$\pm$	3.40E-17	&	1.11E-15	$\pm$	1.60E-17	&	1.13E-16	$\pm$	6.60E-18	&	5.56E-15	$\pm$	6.70E-17	&	14.63	$\pm$	0.88	&	48.19	$\pm$	2.89	\\
NGC 5506	&	3.05E-16	$\pm$	4.90E-17	&	1.48E-15	$\pm$	3.90E-17	&	9.31E-16	$\pm$	2.60E-17	&	1.12E-16	$\pm$	9.20E-18	&	1.40E-16	$\pm$	6.20E-18	&	1.33E-15	$\pm$	1.50E-17	&	10.79	$\pm$	0.65	&	6.45	$\pm$	0.39	\\
NGC 7469	&	4.36E-16	$\pm$	1.30E-16	&	2.56E-15	$\pm$	7.20E-17	&	4.35E-16	$\pm$	1.50E-17	&	4.16E-16	$\pm$	1.10E-17	&	1.67E-16	$\pm$	7.90E-18	&	2.88E-15	$\pm$	1.70E-17	&	30.08	$\pm$	1.80	&	38.19	$\pm$	2.29	\\
IC 4329A	&	1.08E-16	$\pm$	1.20E-17	&	6.16E-16	$\pm$	5.00E-17	&	3.75E-16	$\pm$	1.80E-17	&	3.47E-17	$\pm$	5.00E-18	&	3.73E-17	$\pm$	5.00E-18	&	1.52E-16	$\pm$	1.00E-17	&	2.55	$\pm$	0.15	&	1.32	$\pm$	0.08	\\
Cen A	&	1.13E-15	$\pm$	9.30E-17	&	1.09E-14	$\pm$	1.30E-16	&	3.91E-15	$\pm$	4.40E-17	&	2.00E-15	$\pm$	2.50E-17	&	1.06E-15	$\pm$	1.00E-17	&	2.50E-14	$\pm$	1.00E-16	&	89.54	$\pm$	5.37	&	183	$\pm$	10.98	\\
Circinus	&	3.17E-15	$\pm$	1.10E-16	&	2.27E-14	$\pm$	2.10E-16	&	7.17E-15	$\pm$	6.60E-17	&	3.16E-15	$\pm$	1.80E-17	&	1.63E-15	$\pm$	1.60E-17	&	2.81E-14	$\pm$	1.10E-16	&	346.68	$\pm$	20.80	&	351.23	$\pm$	21.07	\\
NGC 1068	&	6.01E-15	$\pm$	9.90E-17	&	1.26E-14	$\pm$	1.30E-16	&	1.15E-14	$\pm$	8.60E-17	&	5.24E-15	$\pm$	3.70E-17	&	9.70E-16	$\pm$	9.40E-18	&	2.73E-14	$\pm$	1.00E-15	&	211.05	$\pm$	12.66	&	328.18	$\pm$	19.69	\\
NGC 1275	&	3.00E-17			&	1.72E-15	$\pm$	7.20E-17	&	5.48E-17	$\pm$	1.20E-17	&	1.06E-16	$\pm$	8.40E-18	&	1.39E-16	$\pm$	9.00E-18	&	1.49E-15	$\pm$	2.00E-17	&	7.1	$\pm$	0.43	&	6.64	$\pm$	0.40	\\
NGC 1386	&	1.25E-16	$\pm$	2.60E-17	&	3.71E-16	$\pm$	2.20E-17	&	2.61E-16	$\pm$	2.20E-17	&	1.19E-16	$\pm$	1.80E-17	&	4.36E-17	$\pm$	8.00E-18	&	8.01E-16	$\pm$	1.50E-17	&	5.16	$\pm$	0.31	&	9.38	$\pm$	0.56	\\
NGC 7314	&	4.81E-17	$\pm$	1.00E-17	&	2.54E-16	$\pm$	2.20E-17	&	2.14E-16	$\pm$	1.80E-17	&	6.39E-17	$\pm$	8.80E-18	&	2.90E-17	$\pm$	3.00E-18	&	1.04E-15	$\pm$	1.10E-17	&	1.32	$\pm$	0.08	&	5.33	$\pm$	0.32	\\
NGC 7852	&	1.21E-15	$\pm$	9.90E-17	&	2.88E-15	$\pm$	1.20E-16	&	1.89E-15	$\pm$	6.30E-17	&	6.32E-16	$\pm$	1.70E-17	&	2.84E-16	$\pm$	6.40E-18	&	5.18E-15	$\pm$	3.90E-17	&	52.92	$\pm$	3.18	&	56.55	$\pm$	3.39	\\
Mrk 3	&	1.72E-16	$\pm$	2.00E-17	&	1.36E-15	$\pm$	3.60E-17	&	6.79E-16	$\pm$	3.70E-17	&	1.17E-16	$\pm$	1.60E-17	&	9.18E-17	$\pm$	8.00E-18	&	4.66E-16	$\pm$	1.40E-17	&	3.09	$\pm$	0.19	&	2.43	$\pm$	0.15	\\
...              & ...       &    ...  &    ...   &    ...   &  ...   &  ...    &   ...   &   ...  \\
\enddata
\tablecomments{Table~\ref{table:Global measurements} is published in its entirety in the machine-readable format. A portion is shown here for guidance regarding its form and content.}
\end{deluxetable}

\section{Resolved flux measurements for the SHINING galaxies}

\floattable
\begin{deluxetable}{lcccccccccc}
\rotate
\tablecaption{Resolved line fluxes for galaxies in the SHINING sample \label{table:Resolved measurements}}
\tablehead{
\colhead{Source} & \colhead{R.A.} & \colhead{Dec.} & \colhead{\niii} & \colhead{\oi} & \colhead{\oiii} & \colhead{\nii} & \colhead{\oii} & \colhead{\cii~158~$\mu$m} & \colhead{$S_{63}$} & \colhead{$S_{122}$} \\
& J2000 & J2000 & \colhead{W~m$^{-2}$} & \colhead{W~m$^{-2}$} & \colhead{W~m$^{-2}$} & \colhead{W~m$^{-2}$} & \colhead{W~m$^{-2}$} & \colhead{W~m$^{-2}$} &  \colhead{Jy} & \colhead{Jy}\\
}
\startdata  
M 82	&	148.985800	&	69.681435	&		3.19E-16	$\pm$	4.5E-17	&		3.60E-15	$\pm$	2.8E-17	&		1.18E-15	$\pm$	1.7E-17	&		3.59E-16	$\pm$	1.5E-17	&		3.49E-16	$\pm$	3.3E-18	&		4.20E-15	$\pm$	2.7E-17	&		39.14	$\pm$	2.35	&		43.81	$\pm$	2.63	\\
M 82	&	148.979100	&	69.682613	&		2.38E-15	$\pm$	7.6E-17	&		6.85E-15	$\pm$	9.4E-17	&		5.62E-15	$\pm$	6.7E-17	&		1.16E-15	$\pm$	2.6E-17	&		8.19E-16	$\pm$	1.5E-17	&		7.69E-15	$\pm$	5.2E-17	&		120.64	$\pm$	7.24	&		102.69	$\pm$	6.16	\\
M 82	&	148.972715	&	69.684169	&		4.34E-16	$\pm$	2.5E-17	&		2.64E-15	$\pm$	5.1E-17	&		1.62E-15	$\pm$	3.4E-17	&		5.76E-16	$\pm$	1.4E-17	&		3.29E-16	$\pm$	5.6E-18	&		4.22E-15	$\pm$	4.7E-17	&		31.64	$\pm$	1.90	&		31.13	$\pm$	1.87	\\
M 82	&	148.966861	&	69.686261	&	$<$	2.02E-16			&		8.93E-16	$\pm$	2.3E-17	&		5.44E-16	$\pm$	1.9E-17	&		1.59E-16	$\pm$	8.5E-18	&		9.89E-17	$\pm$	4.7E-18	&		1.75E-15	$\pm$	1.6E-17	&		8.69	$\pm$	0.52	&		13.71	$\pm$	0.82	\\
M 82	&	148.957886	&	69.686284	&	$<$	2.02E-16			&		6.16E-16	$\pm$	3.0E-17	&		3.44E-16	$\pm$	2.1E-17	&		1.49E-16	$\pm$	1.2E-17	&		5.72E-17	$\pm$	3.7E-18	&		1.41E-15	$\pm$	1.5E-17	&		6.50	$\pm$	0.39	&		12.58	$\pm$	0.76	\\
M 82	&	148.983381	&	69.679291	&		3.24E-16	$\pm$	6.5E-17	&		3.47E-15	$\pm$	4.2E-17	&		1.05E-15	$\pm$	3.3E-17	&		3.55E-16	$\pm$	1.6E-17	&		3.33E-16	$\pm$	6.0E-18	&		4.66E-15	$\pm$	2.9E-17	&		37.01	$\pm$	2.22	&		39.29	$\pm$	2.36	\\
M 82	&	148.976712	&	69.680425	&		3.79E-15	$\pm$	8.7E-17	&		1.41E-14	$\pm$	1.4E-16	&		8.93E-15	$\pm$	1.5E-16	&		2.04E-15	$\pm$	3.5E-17	&		1.53E-15	$\pm$	3.4E-17	&		1.21E-14	$\pm$	1.3E-16	&		173.00	$\pm$	10.38	&		150.56	$\pm$	9.03	\\
M 82	&	148.969989	&	69.681962	&		3.00E-15	$\pm$	1.1E-16	&		1.04E-14	$\pm$	2.4E-16	&		7.05E-15	$\pm$	1.3E-16	&		1.77E-15	$\pm$	3.9E-17	&		1.10E-15	$\pm$	1.3E-17	&		9.79E-15	$\pm$	7.7E-17	&		109.01	$\pm$	6.54	&		87.14	$\pm$	5.23	\\
M 82	&	148.964291	&	69.684184	&		3.93E-16	$\pm$	3.2E-17	&		2.17E-15	$\pm$	3.7E-17	&		1.26E-15	$\pm$	2.9E-17	&		4.23E-16	$\pm$	7.0E-18	&		2.51E-16	$\pm$	3.7E-18	&		3.79E-15	$\pm$	3.6E-17	&		18.78	$\pm$	1.13	&		16.45	$\pm$	0.99	\\
M 82	&	148.955695	&	69.684269	&		2.43E-16	$\pm$	3.8E-17	&		1.22E-15	$\pm$	2.8E-17	&		6.92E-16	$\pm$	1.7E-17	&		2.32E-16	$\pm$	5.9E-18	&		1.44E-16	$\pm$	3.4E-18	&		2.56E-15	$\pm$	2.0E-17	&		10.70	$\pm$	0.64	&		11.85	$\pm$	0.71	\\
M 82	&	148.980796	&	69.677250	&	$<$	2.02E-16			&		2.27E-15	$\pm$	3.7E-17	&		8.48E-16	$\pm$	2.4E-17	&		3.10E-16	$\pm$	7.6E-18	&		2.05E-16	$\pm$	4.9E-18	&		3.31E-15	$\pm$	2.2E-17	&		19.86	$\pm$	1.19	&		23.73	$\pm$	1.42	\\
M 82	&	148.974088	&	69.678311	&		1.75E-15	$\pm$	4.0E-17	&		8.34E-15	$\pm$	6.4E-17	&		4.52E-15	$\pm$	7.9E-17	&		9.60E-16	$\pm$	2.0E-17	&		7.24E-16	$\pm$	1.7E-17	&		7.65E-15	$\pm$	4.9E-17	&		74.97	$\pm$	4.50	&		66.03	$\pm$	3.96	\\
M 82	&	148.967301	&	69.679779	&		7.77E-15	$\pm$	1.9E-16	&		2.63E-14	$\pm$	5.6E-16	&		1.49E-14	$\pm$	2.5E-16	&		2.96E-15	$\pm$	3.3E-17	&		1.91E-15	$\pm$	2.6E-17	&		1.42E-14	$\pm$	1.1E-16	&		273.31	$\pm$	16.40	&		180.28	$\pm$	10.82	\\
M 82	&	148.961658	&	69.681975	&		2.01E-15	$\pm$	3.6E-17	&		7.47E-15	$\pm$	1.0E-16	&		4.81E-15	$\pm$	7.8E-17	&		1.29E-15	$\pm$	1.4E-17	&		7.62E-16	$\pm$	9.3E-18	&		7.64E-15	$\pm$	7.0E-17	&		62.14	$\pm$	3.73	&		53.09	$\pm$	3.19	\\
M 82	&	148.953143	&	69.682084	&		3.90E-16	$\pm$	2.1E-17	&		2.77E-15	$\pm$	3.5E-17	&		1.27E-15	$\pm$	1.7E-17	&		5.20E-16	$\pm$	1.3E-17	&		3.24E-16	$\pm$	3.9E-18	&		4.22E-15	$\pm$	3.6E-17	&		23.69	$\pm$	1.42	&		25.05	$\pm$	1.50	\\
M 82	&	148.978169	&	69.675194	&	$<$	2.02E-16			&		1.20E-15	$\pm$	1.5E-17	&		5.74E-16	$\pm$	2.1E-17	&		1.81E-16	$\pm$	8.6E-18	&		1.11E-16	$\pm$	3.4E-18	&		1.79E-15	$\pm$	1.8E-17	&		9.98	$\pm$	0.60	&		13.47	$\pm$	0.81	\\
M 82	&	148.971398	&	69.676138	&		5.07E-16	$\pm$	3.8E-17	&		2.80E-15	$\pm$	3.3E-17	&		1.64E-15	$\pm$	2.0E-17	&		4.31E-16	$\pm$	8.1E-18	&		2.52E-16	$\pm$	3.8E-18	&		3.78E-15	$\pm$	2.0E-17	&		24.40	$\pm$	1.46	&		31.43	$\pm$	1.89	\\
M 82	&	148.964646	&	69.677631	&		3.10E-15	$\pm$	4.9E-17	&		1.32E-14	$\pm$	1.7E-16	&		6.61E-15	$\pm$	8.2E-17	&		1.57E-15	$\pm$	3.0E-17	&		9.99E-16	$\pm$	1.3E-17	&		9.59E-15	$\pm$	8.2E-17	&		122.75	$\pm$	7.37	&		88.58	$\pm$	5.31	\\
M 82	&	148.959005	&	69.679759	&		5.71E-15	$\pm$	5.6E-17	&		1.76E-14	$\pm$	1.4E-16	&		1.09E-14	$\pm$	1.1E-16	&		2.63E-15	$\pm$	3.9E-17	&		1.56E-15	$\pm$	1.5E-17	&		1.32E-14	$\pm$	8.1E-17	&		194.73	$\pm$	11.68	&		149.76	$\pm$	8.99	\\
M 82	&	148.950843	&	69.679822	&		4.36E-16	$\pm$	2.4E-17	&		5.03E-15	$\pm$	4.9E-17	&		1.85E-15	$\pm$	2.8E-17	&		5.87E-16	$\pm$	1.6E-17	&		4.31E-16	$\pm$	5.7E-18	&		5.82E-15	$\pm$	3.8E-17	&		51.94	$\pm$	3.12	&		58.47	$\pm$	3.51	\\
M 82	&	148.975263	&	69.673119	&	$<$	2.02E-16			&		6.17E-16	$\pm$	1.4E-17	&		3.26E-16	$\pm$	1.4E-17	&		1.18E-16	$\pm$	4.8E-18	&		5.17E-17	$\pm$	3.0E-18	&		1.00E-15	$\pm$	1.3E-17	&		6.34	$\pm$	0.38	&		9.11	$\pm$	0.55	\\
M 82	&	148.968912	&	69.674144	&	$<$	2.02E-16			&		1.35E-15	$\pm$	2.1E-17	&		6.72E-16	$\pm$	1.4E-17	&		2.30E-16	$\pm$	9.1E-18	&		1.50E-16	$\pm$	3.5E-18	&		2.10E-15	$\pm$	1.7E-17	&		12.51	$\pm$	0.75	&		20.06	$\pm$	1.20	\\
M 82	&	148.962152	&	69.675502	&		4.26E-16	$\pm$	2.5E-17	&		3.53E-15	$\pm$	2.4E-17	&		1.31E-15	$\pm$	1.9E-17	&		4.75E-16	$\pm$	7.5E-18	&		2.92E-16	$\pm$	4.4E-18	&		4.00E-15	$\pm$	4.1E-17	&		28.06	$\pm$	1.68	&		29.82	$\pm$	1.79	\\
M 82	&	148.956434	&	69.677529	&		1.37E-15	$\pm$	2.9E-17	&		7.45E-15	$\pm$	4.5E-17	&		3.07E-15	$\pm$	2.8E-17	&		7.66E-16	$\pm$	1.1E-17	&		5.21E-16	$\pm$	5.8E-18	&		6.35E-15	$\pm$	3.2E-17	&		84.09	$\pm$	5.05	&		63.49	$\pm$	3.81	\\
M 82	&	148.948389	&	69.677554	&	$<$	2.02E-16			&		3.17E-15	$\pm$	2.2E-17	&		7.15E-16	$\pm$	1.5E-17	&		2.36E-16	$\pm$	6.8E-18	&		1.89E-16	$\pm$	4.0E-18	&		3.18E-15	$\pm$	1.9E-17	&		39.28	$\pm$	2.36	&		31.27	$\pm$	1.88	\\
...	&	...	&	...	&	...	&		...	&		...	&		...	&	...	& ...	& ...	& ...	\\
\enddata
\tablecomments{Table~\ref{table:Resolved measurements} is published in its entirety in the machine-readable format. A portion is shown here for guidance regarding its form and content.}
\end{deluxetable}

\bibliography{references.bib}

\end{document}